\documentstyle[preprint,aps,epsf,floats]{revtex}

\newcommand{\nn}{\nonumber}
\newcommand{\ppp}{\mbox{$({\mathbf p'}-{\mathbf p})^2$}}
\newcommand{\spp}{\mbox{$i\bsigma\cdot ({\mathbf p'\times p})$}}
\newcommand{\two}{{\rm 2}}

\def\bmag#1{{|{\mathbf #1}|}}

\def\xslash#1{{\rlap{$#1$}/}}
\def\Dsl{\hbox{/\kern-.6000em D}} 

\def\dsl{\,\raise.15ex\hbox{/}\mkern-13.5mu D}
\def\bsigma{\mbox{\boldmath $\sigma$}}

\def\ms{$\overline{\rm MS}$}
\def\psip#1{\psi_{\mathbf{#1}}}
\def\chip#1{\chi_{\mathbf{#1}}}
\def\bsigma{\mbox{\boldmath $\sigma$}}

\def\abs#1{\left| #1 \right|}
\def\ltap{\ \raise.3ex\hbox{$<$\kern-.75em\lower1ex\hbox{$\sim$}}\ }
\def\gtap{\ \raise.3ex\hbox{$>$\kern-.75em\lower1ex\hbox{$\sim$}}\ }
\def\OMIT#1{}

\begin{document}
\setlength\baselineskip{19pt}

\preprint{\tighten  \hbox{UCSD/PTH 00-04}
}

\title{The QCD heavy-quark potential to order $v^2$: \\
one loop matching conditions}

\author{Aneesh V. Manohar\footnote{amanohar@ucsd.edu} and
Iain W.\ Stewart\footnote{iain@schwinger.ucsd.edu} \\[4pt]}
\address{\tighten Department of Physics, University of California at San
Diego,\\[2pt] 9500 Gilman Drive, La Jolla, CA 92093 }

\maketitle

{\tighten
\begin{abstract}
The one-loop QCD heavy quark potential is computed to order $v^2$ in the color
singlet and octet channels. Several errors in the previous literature are
corrected.  To be consistent with the velocity power counting, the full
dependence on $|{\bf p}+{\bf p'}|/|{\bf p'}-{\bf p}|$ is kept. The matching
conditions for the NRQCD one-loop potential are computed by comparing the QCD
calculation with that in the effective theory. The graphs in the effective
theory are also compared to terms from the hard, soft, potential, and ultrasoft
regimes in the threshold expansion.  The issue of off-shell versus on-shell
matching and gauge dependence is discussed in detail for the
$1/(m\!\abs{\mathbf k})$ term in the potential. Matching on-shell gives a
$1/(m\!\abs{\mathbf k})$ potential that is gauge independent and does not
vanish for QED.
\end{abstract}
\pacs{12.39.Hg,11.10.St,12.38.Bx}
}
\vspace{0.7in}

\newpage 

\section{Introduction}

For processes involving a non-relativistic heavy quark and antiquark, it is
useful to combine the $\alpha_s$ expansion of QCD with an expansion in powers
of the relative velocity $v$. The scattering of a heavy quark and antiquark,
$Q({\mathbf p}) + {\bar Q} ({-\mathbf p}) \to  Q({\mathbf p}^\prime) + {\bar Q}
({-\mathbf p}^\prime)$, can be described using a potential $V$, which has an
expansion in powers of the velocity $v$ and $\alpha_s$,
\begin{eqnarray} \label{V1}
  & & V({\bf p},{\bf p'}) = \sum_{n=-2}^{\infty} V^{(n)} , \qquad
     V^{(n)} = \sum_{j=1}^\infty  V^{(n,j)} , \nn\\[5pt]
  & & \mbox{where}\quad V^{(n)}\sim  v^n \,, \qquad\quad
     V^{(n,j)}\sim  v^n \alpha_s^j  \,.
\end{eqnarray}
An important complication of non-relativistic scattering is the presence of two
low-energy scales, $mv$ and $mv^2$, which correspond to the momentum and energy
of the heavy quark, respectively. In this paper, we will assume that $mv^2 \gg
\Lambda_{\rm QCD}$, and ignore any non-perturbative effects.

The first term in Eq.~(\ref{V1}), $V^{(-2,1)}$, is the static Coulomb
potential  at tree-level,
\begin{eqnarray}
  V^{(-2,1)} = C{4 \pi \alpha_s\over \abs{{\mathbf p} - {\mathbf p}^\prime}^2}
  \,,
\end{eqnarray}
where the color factor $C$ depends on the relative color state of the incident
quark and antiquark. For $j>1$, $V^{(-2,j)}$ are perturbative corrections to
the Coulomb potential  which are known to two loops~\cite{Peter,Schroder}.

The static potential for a color singlet $Q \bar Q$ pair is  often defined in
terms of the expectation value of a rectangular Wilson loop of width $R$ and
length $T$:
\begin{eqnarray} \label{VWilson}
  V^{(-2)}(R) &=& - \lim_{T\to \infty} \frac{1}{iT} \ln\bigg\langle {\rm Tr}\:
   P\, \exp\Big( i g \oint dx_\mu A_a^\mu T^a \Big) \bigg\rangle \,.
\end{eqnarray}
The Feynman diagrams corresponding to this expectation value build up the
exponential of the static potential.  As a result, in computing the static
potential $V^{(-2,j)}$ at order $j$, iterations of lower order terms in the
potential, $V^{(-2,n)},\ n<j$, are subtracted. At three loops Appelquist, Dine
and Muzinich~\cite{adm} have shown that infrared divergences are encountered in
$V^{(-2,4)}(R)$ which are not canceled by simply subtracting iterations of the
lower order potentials. Thus, using the definition in Eq.~(\ref{VWilson}) at
higher orders in perturbation theory, or generalizing this approach to
subleading terms in the $v$ expansion, becomes cumbersome as the set of
subtractions become more complicated, and perturbative subtractions are
insufficient to render the potential in Eq.~(\ref{VWilson}) well-defined.

A convenient framework for investigating the expansion in Eq.~(\ref{V1}) is the
effective field theory for non-relativistic QCD (NRQCD), formulated with a
power counting in
$v$~\cite{Caswell,BBL,Labelle,lm,Manohar,gr,ls,Pineda,P2,Beneke,Gries,P3,LMR}.
In the effective field theory, it is more convenient to define the potential as
the Wilson coefficient of a four-quark operator~\cite{Pineda}
\begin{eqnarray} \label{Lp0}
{\mathcal L}_p= - \sum_{\mathbf p,p'} V_{\alpha\beta\lambda\tau} \!
  \left({\bf p} ,{\bf p^\prime}\right)\ \mu^{2\epsilon} \:
  \Big[ {\psip {p^\prime}}_\alpha^\dagger(x)\:
  {\psip p}_\beta(x)\: {\chip {-p^\prime}}_\lambda^\dagger(x) \:
  {\chip {-p}}{}_\tau(x) \Big] \,.
\end{eqnarray}
In Eq.~(\ref{Lp}), the fields $\psip {}$ and $\chip {}$ annihilate a quark and
an antiquark, respectively. The fields are labelled by a momentum ${\bf p}$, and
a greek index for their color and spin. The operator in Eq.~(\ref{Lp0}) is
local on the scale $x\sim 1/mv^2$, but non-local on the scale ${\mathbf p}\sim
mv$.  The potential $V$ is computed as a matching coefficient at the scale
$\mu=m$ between QCD and an effective theory for non-relativistic QCD valid
below the scale $m$. The effective theory is constructed to have the same
infrared structure as perturbative QCD for the two heavy quark system.
Therefore, defining the potential as a matching coefficient provides an
infrared safe definition. For instance, Ref.~\cite{psIR} shows how the
three-loop matching potential $V$ is infrared safe, despite the divergence in
the QCD potential of Appelquist, Dine and Muzinich\cite{adm}.

Although several different formulations of the effective theory for
non-relativistic QCD are currently in
use~\cite{BBL,Labelle,lm,Manohar,gr,ls,Pineda,P2,Beneke,Gries,P3,LMR,kp,Brambilla,amis},
certain universal features have emerged.  The on-shell degrees of freedom in
the effective theory include quarks with energy $E\sim mv^2$ and momentum
$p\sim mv$, soft gluons with $E\sim p\sim mv$, and ultrasoft gluons with $E\sim
p\sim mv^2$.  The soft and ultrasoft modes are distinct, for instance a
consistent power counting in $v$ demands that the ultrasoft gluon interactions
are multipole expanded\cite{Labelle,gr}, while soft gluon interactions are not.
The soft gluons are essential to correctly reproduce the beta function in the
effective theory~\cite{Gries}, and run between the scales $m$ and $mv$. Other
massless on-shell fields, such as light quarks, will have ultrasoft and soft
components too.  There are also important off-shell field components, such as
the exchange of gluons with $E\sim mv^2$ and $p\sim mv$ that build up the
potential.  Soft heavy quarks with $E\sim p\sim mv$ are also off-shell relative
to the heavy quark states of interest. These off-shell field components can be
integrated out of the Lagrangian in the effective theory. Doing this leaves a
Lagrangian that is non-local at the scale $mv$, but local at $mv^2$.  This
procedure, which treats the potential components as four quark operators, was
first seriously investigated in Ref.~\cite{Pineda}, and the resulting effective
theory is referred to as pNRQCD.  In Ref.~\cite{Pineda} it was proposed that the
matching onto effective theories should take place in two stages: at $\mu=m$ one
matches QCD onto NRQCD as originally defined in Ref.~\cite{BBL}, and then
matches NRQCD onto pNRQCD at the scale $\mu=m v$.

The matching of four-quark operators at $m$ was considered in Ref.~\cite{P2},
following the proposal in Ref.~\cite{Manohar} that the matching procedure should
be similar to that in HQET. However, in general this procedure seems to be
slightly problematic\cite{LMR}.  First note that it is necessary to include the
kinetic term for the potential quarks immediately below the scale $m$.  For
instance, this is necessary to reproduce the threshold value of the two-loop
anomalous dimension for the heavy quark production current \cite{prodcurrent} in
the effective theory \cite{LMR}.  With the kinetic energy term included at
leading order, the consistency of the $v$ power counting forces us to have both
ultrasoft and soft modes {\em right below} $m$ (along with the multipole
expansion etc.).  If we note that the kinetic energy term is necessary to
correctly reproduce the Coulombic infrared divergences in the effective theory,
then it might seem obvious that it must be included in order to have the correct
infrared structure.  At one-loop, matching exactly at threshold in dimensional
regularization seemed to provide a method of avoiding this~\cite{P2}, but
this procedure fails at two-loops.

Since we must include the kinetic energy term for all $\mu<m$ it seems quite
natural to immediately consider matching at $\mu=m$ onto an effective theory
where the off-shell potential gluons and soft quarks have been integrated out.
Such a formulation was proposed in Ref.~\cite{LMR} and will be considered in
this paper. It will be referred to as vNRQCD. In this theory it is more
appropriate to consider the running from $m$ to the scales $mv$ and $mv^2$ in a
single step. This is accomplished by using the velocity renormalization
group~\cite{LMR}. Once the vNRQCD effective theory has been run down to the low
scale the soft degrees of freedom have served their purpose and may be
integrated out.  In this final effective theory no additional running needs to
be considered.

In this paper we compute the one-loop matching for the $Q\bar Q$ and $QQ$
potentials between QCD and vNRQCD up to corrections of order $v^2$.  Ours
results are compared to the matching calculation of Pineda and Soto~\cite{P2}
for the four-quark operators and to the threshold expansion~\cite{Beneke}. The
main difference between the the non-relativistic theories in Refs.~\cite{LMR}
and \cite{Pineda,P2,Brambilla} is the way in which large logarithms of $v$ are
summed in the effective theory.  There is also some difference in the matching
corrections; some contributions in vNRQCD that arise at the scale $\mu=m$ are
instead computed at $\mu=mv$ in pNRQCD. We will discuss the differences between
the two approaches in the text of the paper.

The full QCD calculation of the $Q \bar Q$ scattering to order $\alpha_s^2 v^0$
in the non-relativistic expansion has been done before~\cite{Gupta,Yndurain}.
In calculating the potentials, Ref.~\cite{Gupta} performs an  additional
expansion, assuming
\begin{eqnarray} \label{ratio}
 {  ({\bf p}'\,+{\bf p}\,)^2 \over ({\bf p}\,'\,-{\bf p}\,)^2 } \ll 1 \,.
\end{eqnarray}
In the usual $v$ power counting, $\mathbf p$ and $\mathbf p^\prime$ are both of
order $v$, so the ratio in Eq.~(\ref{ratio}) is of order unity, and cannot be
treated as small. For this reason, in section~II we redo the QCD calculation
keeping the full dependence on this ratio for both the color singlet and octet
channels. For the order $v^0$ spin-independent potential, terms proportional to
$({\bf p}+{\bf p'})^2$ which were dropped in previous calculations
\cite{Gupta,Yndurain} are necessary to match infrared divergences between the
full and effective theories.

In section III we discuss the order  $v^0$ potential of the form
\begin{eqnarray} \label{Dptnl}
 \frac{ ({\bf p'}^2-{\bf p}^2)^2 }{ m^2 ({\bf p}\,'\,-{\bf p}\,)^4 } \,.
\end{eqnarray}
For free quark states this potential vanishes on-shell by energy conservation,
${\bf p'}^2 = {\bf p}^2$, and need not be included in the potential if one uses
on-shell matching~\cite{Georgi}. The potential in Eq.~(\ref{Dptnl}) gives a
non-zero contribution in loops or in matrix elements with external Coulomb
states, and is often included in the quark potential. For instance, the usual
Breit Hamiltonian includes a potential of the form in Eq.~(\ref{Dptnl}). A
potential that is only non-zero off-shell can be gauge dependent, and including
the potential of Eq.~(\ref{Dptnl}) induces  gauge dependence in the
coefficients of on-shell potentials.  Using an off-shell potential gives
correct results if all calculations are performed in  the same gauge. We will
use an on-shell basis for the potential where the term in Eq.~(\ref{Dptnl})
does not occur. The matching coefficients for the on-shell potential are gauge
independent, since scattering amplitudes are measurable quantities.  The
difference between the on-shell matching potential and the Breit Hamiltonian is
compensated for by a corresponding change in the $1/|{\bf p'}-{\bf p}|$
potential, as discussed in more detail in section~III (see also
Refs.~\cite{melnikov,hoang,mm1bram,hasebe,faustov}).

In section~IV we extend all of our results for the quark-antiquark potential to
the quark-quark potential and in section~V we give the QED limit of our results.
Section VI gives our conclusions.

\section{Matching the potential to ${\cal O}$\lowercase{$(v^2)$} }

The vNRQCD effective Lagrangian has the form
\begin{eqnarray}
 {\cal L} = {\cal L}_u + {\cal L}_s + {\cal L}_p \,.
\end{eqnarray}
The ultrasoft Lagrangian ${\cal L}_u$ involves the fields $\psip p$ which
annihilate a quark, $\chip p$ which annihilate an antiquark, and $A^\mu$ which
annihilate and create ultrasoft gluons.  The potential Lagrangian ${\cal L}_p$
contains operators with four or more quark fields including the quark-antiquark
potential. Finally the soft Lagrangian ${\cal L}_s$ contains all terms that
involve soft fields which have energy and momenta of order $mv$. Heavy quarks
with soft energy and momenta are off-shell and are therefore integrated out, so
they do not appear explicitly in ${\cal L}_s$.  The terms we need in the
ultrasoft Lagrangian are
\begin{eqnarray} \label{Lu}
  {\mathcal L}_u &=& -{1\over 4}F^{\mu\nu}F_{\mu \nu} + \sum_{\mathbf p}
   \psip p ^\dagger   \Biggl\{ i D^0 - {\left({\bf p}-i{\bf D}\right)^2
   \over 2 m} +\frac{{\mathbf p}^4}{8m^3} \Biggr\} \psip p  \nn\\
   & & + \chip p ^\dagger
   \Biggl\{ i D^0 - {\left({\bf p}-i{\bf D}\right)^2
   \over 2 m} +\frac{{\mathbf p}^4}{8m^3} \Biggr\} \chip p \,.
\end{eqnarray}
The covariant derivative on $\psip p$ and $\chip p$ contain the color matrices
$T^A$ and $\bar T^A$ for the $\bf 3$ and $\bf {\bar 3}$ representations,
respectively.  Here $D^\mu$ involves only the ultrasoft gluon fields: $D^\mu =
\partial^\mu + i g \mu_U^\epsilon A^\mu=(D^0,-\mathbf{D})$, so
$D^0=\partial^0+ig\mu_U^\epsilon A^0$, ${\mathbf D}={\mathbf
\nabla}-ig\mu_U^\epsilon {\mathbf A}$.  The factors of the ultrasoft scale
parameter $\mu_U^\epsilon$ are included to make $g=g(\mu_U)$ dimensionless in
$d=4-2\epsilon$ dimensions.  The ultrasoft gluon field $A^\mu$ is order $v^2$
and has dimension $1-\epsilon$, so $A^\mu \sim (m v^2)^{1-\epsilon}$.
Consistency of the $v$ power counting for the covariant derivative in $d$
dimensions then requires $\mu_U=m\nu^2$ where $\nu\sim v$. This reproduces the
dependence of $\mu_U$ on the subtraction velocity $\nu$ given in
Ref.~\cite{LMR}.  The terms we need in the soft Lagrangian are
\begin{eqnarray}\label{Lsoft}
 {\mathcal L}_s &=& \sum_{q} \bigg\{ \abs{q^\mu A^\nu_q -
 q^\nu A^\mu_q}^2 + \bar \varphi_q\, \xslash{q}\, \varphi_q  +
 \bar c_q \, q^2 c_q \ \bigg\}  \\
 && - g^2 \mu_S^{2\epsilon} \sum_{{\mathbf p,p'},q,q'} \bigg\{ \frac12\, {\psip
 {p^\prime}}^\dagger\:
 [A^\mu_{q'},A^\nu_{q}] U_{\mu\nu}^{(\sigma)}\: {\psip p}\: + \frac12\,
 {\psip {p^\prime}}^\dagger\: \{A^\mu_{q'},A^\nu_{q}\} W_{\mu\nu}^{(\sigma)}\:
 {\psip p}\: \nn \\
 && + {\psip {p^\prime}}^\dagger\: [\bar c_{q'},c_{q}] Y^{(\sigma)}\:
 {\psip p}\: + ({\psip {p^\prime}}^\dagger\: T^B Z_\mu^{(\sigma)}\:
 {\psip p} ) \:(\bar \varphi_{q'} \gamma^\mu T^B \varphi_q)  \bigg\}
 + (\psi \to \chi,\: T\to \bar T) \,, \nn
\end{eqnarray}
where $A_q$, $c_q$, and $\varphi_q$ are soft gluons, ghosts, and massless
quarks. The functions $U$, $W$, $Y$, and $Z$ are given in
Appendix~\ref{App_soft}. After integrating out the soft quarks the Lagrangian
${\cal L}_s$ is no longer manifestly gauge invariant with respect to gauge
transformations at the scale $m v$. Therefore, determining the dependence of
the soft scale parameter $\mu_S$ on $\nu$ may seem more difficult than the
ultrasoft case. However, prior to integrating out the soft quarks the
combination $g \mu_S^\epsilon A_q$ is from a covariant derivative, and
$A_q\sim (m v)^{1-\epsilon}$, yielding $\mu_S \sim m\nu$ in agreement with
Ref.~\cite{LMR}. In Eq.~(\ref{Lsoft}), $g=g(\mu_S)$. In general it is important
to realize that the $v$ scaling of $\mu_U$ and $\mu_S$ are different.  If the
matching calculation is performed at the scale\footnote{\tighten It is not
necessary to match exactly at $m$. If the matching scale is $\mu=\mu_h\sim m$,
then one still sets $\mu_U=\mu_S=\mu_h$ and $\nu=1$, and factors of
$\ln(\mu_h/m)$ appear in the matching coefficients. For convenience we choose
$\mu_h=m$ in this paper.} $m$, then for this computation $\mu=\mu_S=\mu_U=m$ and
$\nu=1$ (where the usual QCD scale parameter is denoted by $\mu$). Therefore, for
the matching at $m$ it is not essential to distinguish between $\mu_S$ and
$\mu_U$.

The potential interaction relevant for our calculation is
\begin{eqnarray}\label{Lp}
{\mathcal L}_p= - \sum_{\mathbf p,p'} V_{\alpha\beta\lambda\tau}
  \left({\bf p},{\bf p^\prime}\right)\ \mu_S^{2\epsilon}\:
  {\psip {p^\prime}}_\alpha^\dagger\: {\psip p}_\beta\:
  {\chip {-p^\prime}}_\lambda^\dagger\:  {\chip {-p}}{}_\tau .
\end{eqnarray}
The factor of $\mu_S^{2\epsilon}$ is included so that in $d=4-2\epsilon$
dimensions the potential $V$ has dimension $-2$. $\alpha, \beta, \lambda, \tau$
denote color and spin indices.  We will use the basis in which the potential is
written as a linear combination of $1 \otimes 1$ and $T^a \otimes \bar T^a$ in
color space.  One can convert to the color singlet and octet potential using
the linear transformation
\begin{eqnarray} \label{QQbarProj}
 \left[\begin{array}{c} V_{\rm singlet} \cr V_{\rm octet} \end{array}\right]
 =\left[\begin{array}{ccc} 1 &  & -C_F \cr
    1 &  & {1\over 2} C_A - C_F \cr
 \end{array}\right]
 \left[\begin{array}{c} V_{1\otimes 1} \cr V_{T\otimes \bar T} \end{array}
 \right]\,,
\end{eqnarray}
where $C_F=(N_c^2-1)/(2N_c)$ and $C_A=N_c$. We will also need the invariants
$C_1=(N_c^2-1)/(4N_c^2)$ and $C_d=N_c-4/N_c$. These arise in the identities
\begin{eqnarray} \label{TTiden}
  T^A T^B \otimes \bar T^A \bar T^B &=& - \frac14 (C_A + C_d) T^A \otimes \bar
    T^A + C_1 1 \otimes 1 \,, \nn \\
  T^A T^B \otimes \bar T^B \bar T^A &=& \phantom{-} \frac14
    (C_A - C_d) T^A \otimes \bar T^A + C_1 1 \otimes 1 \,.
\end{eqnarray}
Written as a matrix, the order $v^{-2}$ Coulomb potential is
\begin{eqnarray}
 V^{(-2)} &=&  (T^A \otimes \bar T^A) { {\cal V}_c^{(T)} \over {\mathbf k}^2}
  + (1 \otimes 1) { {\cal V}_c^{(1)} \over {\mathbf k}^2} \,,
\end{eqnarray}
where ${\bf k} = {\bf p'} - {\bf p}$, and the coefficients ${\cal V}_c^{(T)}$
and ${\cal V}_c^{(1)}$ have an expansion in $\alpha_s$.

The order $v^0$ potential includes
\begin{eqnarray} \label{V0}
 V^{(0)} &=&  (T^A \otimes \bar T^A) \left[ { {\cal V}_2^{(T)} \over m^2 }
 + { {\cal V}_r^{(T)}\: ({\mathbf p^2 + p^{\prime 2}}) \over 2\, m^2\,
 {\mathbf k}^2}
 + { {\cal V}_s^{(T)} \over m^2}\, {\mathbf S}^2 + {{\cal V}_\Lambda^{(T)}
 \over m^2}\, \Lambda({\mathbf p^\prime ,p}) + { {\cal V}_t^{(T)} \over  m^2}\,
 {\bf T}({\mathbf k})\right] \nn \\
 && + (1\otimes 1) \left[ { {\cal V}_2^{(1)} \over m^2 } +
 { {\cal V}_s^{(1)} \over m^2}\: {\mathbf S}^2
 \right] \,,
\end{eqnarray}
where
\begin{eqnarray}
\mathbf S &=& { {\mathbf \bsigma_1 + \bsigma_2} \over 2},
 \qquad \Lambda({\mathbf p^\prime, p }) = -i {\mathbf S \cdot ( p^\prime
 \times p) \over  {\mathbf k}^2 },\qquad
 {\bf T}({\mathbf k}) = {\mathbf \bsigma_1 \cdot \bsigma_2} - {3\, {\mathbf k
 \cdot \bsigma_1}\,  {\mathbf k \cdot \bsigma_2} \over {\mathbf k}^2} \,,
\end{eqnarray}
and $\bsigma_1/2$ and $\bsigma_2/2$ are the spin-operators on the quark and
anti-quark.  Note that on-shell ${\bf p'\,^2=p^2}$, but we have written
${\mathbf p^2 + p^{\prime 2}}$ in Eq.~(\ref{V0}) so that the Lagrangian  ${\cal
L}_p$ is hermitian.
\begin{figure}
  \epsfxsize=8cm \hfil\epsfbox{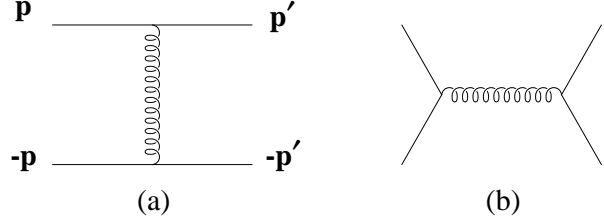}\hfill \\*[-15pt]
{\tighten \caption{QCD diagrams for tree level matching.} \label{fig_tree} }
\end{figure}
The tree level diagram in Fig.~\ref{fig_tree}a generates terms of ${\cal
O}(v^{2k} \alpha_s),\ k\ge-1$, in the QCD potential. Matching at $\mu=m$,
$\nu=1$ gives
\begin{eqnarray}  \label{bc}
 {\cal V}_c^{(T)} &=& \ \ \ 4 \pi \alpha_s(m)\,, \qquad
 {\cal V}_c^{(1)} = 0 \,, \qquad\qquad\qquad
 {\cal V}_r^{(T)} = 4 \pi \alpha_s(m)\,, \nn\\
 {\cal V}_s^{(T)} &=& -\frac{4 \pi \alpha_s(m)}{3}\,, \qquad
 {\cal V}_\Lambda^{(T)} = -6 \pi \alpha_s(m) \,,\qquad
 {\cal V}_t^{(T)} = -\frac{\pi \alpha_s(m)}{3} \,, \nn\\
 {\cal V}_s^{(1)} &=&  0 \,, \qquad\qquad\qquad\
 {\cal V}_2^{(T)} = 0 \,,\qquad\qquad\qquad\
 {\cal V}_2^{(1)} = 0 \,.
\end{eqnarray}

The annihilation diagram in Fig.~\ref{fig_tree}b generates terms of order
$\alpha_s v^{2k}, k \ge 0$ in the potential.  Using Fierz  identities and
charge conjugation, these operators can be transformed into the  basis in
Eq.~(\ref{V0}) and give additional contributions to the matching. Only ${\cal
V}_{s}^{(T,1)}$ receive non-zero annihilation contributions at  tree level:
\begin{eqnarray} \label{bc2}
 {\cal V}_{s,a}^{(T)} = {1 \over N_c}\: \pi\, \alpha_s(m) \,, \qquad
 {\cal V}_{s,a}^{(1)} =  {(N_c^2-1)\over 2N_c^2}\: \pi\, \alpha_s(m) \,.
\end{eqnarray}
The complete tree level matching is given by adding the terms in
Eqs.~(\ref{bc}) and (\ref{bc2}). We have found it convenient to  distinguish
the annihilation contributions by including an additional subscript $a$ on the
coefficients they generate.  The leading-log values of the order  $v^0$
potentials in Eq.~(\ref{V0}) were calculated in Refs.~\cite{chen,amis}, but are
not needed here. Nonzero values for ${\cal V}_2^{(T,1)}$ are generated in the
renormalization group flow below the scale $m$~\cite{amis}, as well as by the
one-loop matching as we will see below.

At one-loop the matching onto QCD gives order $1/v$ terms of the form
\begin{eqnarray} \label{Lk}
 V^{(-1)} &=& { \pi^2 \over m\,|{\mathbf k}| }\,
  \Big[ {\cal V}_k^{(T)} (T^A \otimes \bar T^A) +
  {\cal V}_k^{(1)} (1\otimes 1)\: \Big]  \,,
\end{eqnarray}
where the coefficients ${\cal V}_k^{(T,1)}$ are dimensionless. In
$d=4-2\epsilon$ dimensions the one-loop matching produces a potential with the
dependence $\mu^{2\epsilon}/|{\bf k}|^{1+2\epsilon}$. We have chosen to define
$V^{(-1)}$ by taking the $d\to 4$ limit, which differs from the definition of
this operator used in Ref.~\cite{BSS}.

\subsection{The QCD Calculation}

To perform the potential matching calculation, we consider the on-shell $Q \bar
Q$ scattering amplitude in QCD and in vNRQCD to order $\alpha_s^2 v^0$.  We
will use Feynman gauge, regulate infrared divergences with a finite gluon
mass\footnote{\tighten Using a finite gluon mass is dangerous in the
presence of diagrams with the non-abelian gluon vertices. All such diagrams we
require here are IR finite in Feynman gauge.} $\lambda$, and renormalize
ultraviolet divergences with dimensional regularization and the \ms\ scheme.
Since the calculation is performed on-shell the resulting matching coefficients
will be gauge independent.

We begin by considering the QCD diagrams. The most complicated diagram is the
QCD box diagram~\cite{Redhead,Nieuw} which includes contributions of order
$1/v^3$, $1/v^2$, $1/v$ and $v^0$, as well as higher order terms which we do
not need in this paper.
\begin{eqnarray} \label{box}
\begin{picture}(65,20)(1,1)
 \epsfxsize=2.cm \lower12pt \hbox{\epsfbox{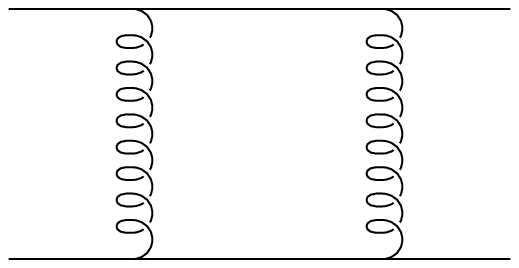}}
\end{picture}
 &=& \frac{i \alpha_s^2}{k^2}\, (T^A T^B \otimes \bar T^A \bar T^B) \Bigg[
  { 2 i m \pi  \over  \: p} \ln\Big( \frac{k^2} {
  \lambda^2} \Big) - 4 \ln\Big( \frac{k^2}{\lambda^2} \Big) \\
  &+& \frac{\pi^2 k}{m p\, (2p+k)} \bigg\{ 6p^2\!+\!\frac{5 p k}{2}\!+\!
  \frac{k^2}{4} -3 k^2 \Lambda - \Big(p k +\frac{5 k^2}{6}\Big){\bf S}^2
  -\frac{k^2 {\bf T}}{12} -\frac{k {\bf R}}{4(2p+k)} \bigg\}\nn\\[5pt]
  &+& \frac{6 i\pi k^2 \Lambda }{m p\,(4p^2-k^2)}\bigg\{ k^2 \ln\Big(
  \frac{2p}{\lambda}\Big)-4 p^2 \ln\Big(\frac{k}{\lambda}\Big) \bigg\} \nn\\
  &+& \frac{i \pi k^2 {\bf S}^2}{3 m p\, (4 p^2-k^2) } \bigg\{ k^2-4 p^2
  +(5k^2-12p^2) \ln\Big(\frac{2p}{\lambda}\Big)-(4 p^2 +k^2)
  \ln\Big(\frac{k}{\lambda}\Big) \bigg\} \nn\\
  &+& \frac{i \pi k^2 {\bf T}}{6 m p\, (4 p^2-k^2) } \bigg\{ 4 p^2 -k^2
  +k^2 \ln\Big(\frac{2p}{\lambda}\Big)+(-8 p^2 +k^2) \ln\Big(\frac{k}
  {\lambda}\Big) \bigg\}  \nn \\
  &-& \frac{i \pi k^2 {\bf R}}{ 2 m p\, (4 p^2-k^2)^2 } \bigg\{ k^2-4 p^2 +
   (4 p^2+k^2) \ln\Big( \frac{2p}{k}\Big) \bigg\}\nn\\
 &+& \frac{i \pi }{2 m p\, (4 p^2-k^2) } \bigg\{ 4 p^2 k^2-k^4-k^2(4p^2+k^2)
  \ln\Big(\frac{2p}{\lambda}\Big) \nn \\
 &+& (80p^4-16 p^2 k^2+k^4 )\ln\Big(\frac{k}{\lambda}\Big) \bigg\}
 -\frac{6k^2}{m^2}+\frac{8k^2 {\bf S^2}}{3 m^2} -\frac{k^2 {\bf T}}{3 m^2} \nn\\
 &+& \ln\Big(\frac{\lambda}{k}\Big) \bigg\{ \frac{56 p^2}{3 m^2}
 -\frac{12k^2 \Lambda}{m^2} -\frac{8 k^2 {\bf S}^2}{3 m^2}-\frac{2k^2 {\bf T}}
 {3 m^2} \bigg\} +\ln\Big(\frac{k}{m}\Big) \bigg\{ \frac{4k^2}{m^2}-
 \frac{2k^2 {\bf S}^2}{m^2} \bigg\} \Bigg]  \,, \nn
\end{eqnarray}
where $k=|{\mathbf k}|$, $p=|{\mathbf p}|=|{\mathbf p^\prime}|$ and
\begin{equation}
  {\bf R} = ({\mathbf p+p'})\! \cdot \bsigma_1\, ({\mathbf p+p'})\! \cdot
   \bsigma_2 \,. \label{Rdef}
\end{equation}
The real part of the order $1/v$ amplitude agrees with Ref.~\cite{Gupta} in the
limit $p \to k/2$.  We have kept the full $p$ dependence since taking this
limit is not justified by the power counting. We have also kept imaginary terms
generated by the cut amplitude to emphasize how these terms are correctly
reproduced in the effective theory. The real part of the spin dependent order
$v^0$ amplitude agree with the result in Ref.~\cite{PTN}. The real part of the
spin independent order $v^0$ amplitude agrees with Ref.~\cite{Yndurain}, except
for the order $v^0$ $\ln(k)$ and $\ln(\lambda)$ dependence. The difference is
due to the condition $|{\bf p'+p}| \ll k$ which was imposed in
Refs.~\cite{Gupta,Yndurain}, but which violates the $v$ power counting. (For the
crossed box given below in Eq.~(\ref{qcdother}a), the order $v^0$ $\ln(\lambda)$
and $\ln(k)$ also differs from Ref.~\cite{Yndurain}.) The remaining direct
scattering diagrams are less complicated since they have no cuts:
\begin{mathletters} \label{qcdother}
\begin{eqnarray}
\begin{picture}(65,20)(1,1)
 \epsfxsize=2.cm \lower12pt \hbox{\epsfbox{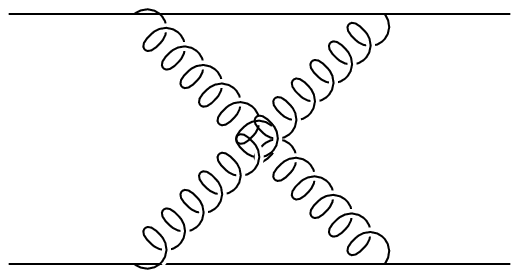}}
\end{picture}
 &=& \frac{i \alpha_s^2}{k^2}\, (T^A T^B \otimes \bar T^B \bar T^A) \Bigg[
  4 \ln\Big( \frac{k^2}{\lambda^2} \Big)
  -\frac{\pi^2 {k}}{m}\ \\*
  && + \ln\Big(\frac{\lambda}{k}\Big) \bigg\{ \frac{8k^2}{3 m^2}
  -\frac{56 p^2}{3 m^2} +\frac{12k^2\Lambda}{m^2}
  +\frac{8 k^2 {\bf S}^2}{3 m^2}+ \frac{2 k^2 {\bf T}}{3 m^2} \bigg\} \nn \\
  &&+\ln\Big(\frac{k}{m}\Big) \bigg\{ -\frac{6 k^2}{m^2}+\frac{2 k^2 {\bf S}^2}{m^2}
  \bigg\} + \frac{2 k^2}{m^2}-\frac{2k^2 {\bf S}^2}{3 m^2}+\frac{k^2 {\bf T}}
  {3 m^2} \Bigg] \,, \nn \\[5pt]
\begin{picture}(65,20)(1,1)
 \epsfxsize=2.cm \lower12pt \hbox{\epsfbox{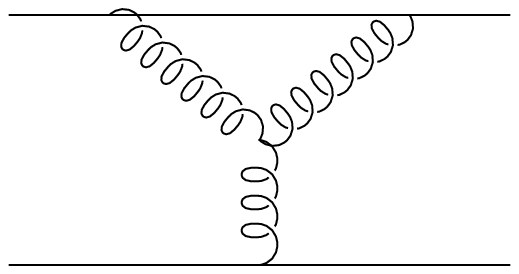}}
\end{picture}
+
\begin{picture}(65,20)(1,1)
 \epsfxsize=2.cm \lower12pt \hbox{\epsfbox{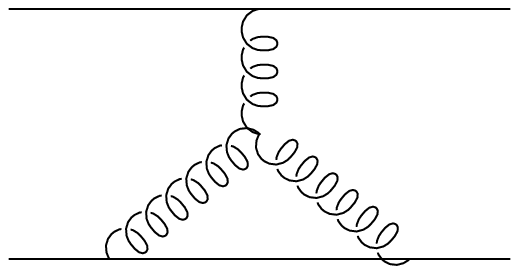}}
\end{picture}
 &=& -\frac{i \alpha_s^2}{k^2}\, C_A (T^A \otimes \bar T^A) \Bigg[3  \ln\Big(
 \frac{\mu^2}{m^2} \Big)+4 +\frac{\pi^2 {k}}{2 m} \nn \\
 && + \ln\Big(\frac{k}{m}\Big) \bigg\{\frac{4 k^2}{m^2} -\frac{4 k^2 {\bf S}^2}{3m^2} -
 \frac{k^2 {\bf T} }{3m^2} - \frac{4 k^2 \Lambda}{m^2} \bigg\} \nn\\
 && + \ln\Big(\frac{\mu}{m}\Big) \bigg\{
  \frac{6 p^2 }{m^2} -\frac{9 k^2 \Lambda}{m^2} -\frac{2 k^2 {\bf S}^2}
  { m^2}-\frac{k^2 {\bf T}}{2 m^2} \bigg\} \nn\\
 && + \frac{4 p^2 }{m^2}-\frac{8 k^2 \Lambda}{m^2} - \frac{2 k^2 {\bf S^2}}
  {m^2}-\frac{k^2 {\bf T}}{2m^2} \Bigg]  \,,\\[15pt]
\begin{picture}(65,20)(1,1)
 \epsfxsize=2.cm \lower8pt \hbox{\epsfbox{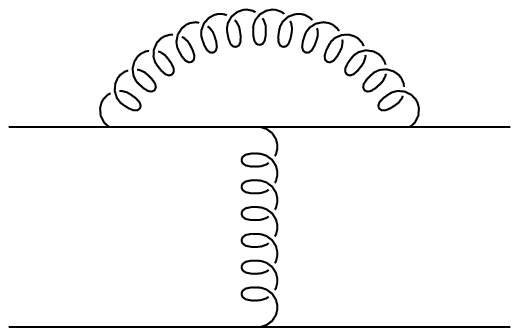}}
\end{picture}
+\
\begin{picture}(65,20)(1,1)
 \epsfxsize=2.cm \lower20pt \hbox{\epsfbox{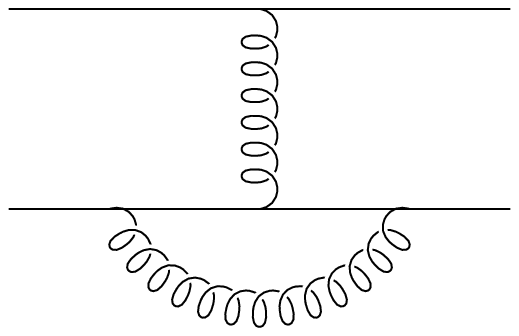}}
\end{picture}
 &=& -\frac{ i \alpha_s^2}{k^2}\, (2C_F\!-C_A) (T^A \otimes \bar T^A)
  \Bigg[ \bigg\{ 2 \ln\Big( \frac{\lambda^2}{m^2} \Big) \!+\!
  \ln\Big( \frac{\mu^2}{m^2} \Big)\!+\!4 \bigg\} M^0 \nn\\
  && +\frac{k^2}{m^2} -\frac{2 k^2 {\bf S^2}}{3 m^2} -\frac{k^2 {\bf T}}{6m^2}
  -\frac{2 k^2 \Lambda}{m^2} +\frac{4 k^2}{3 m^2}\ln\Big(\frac{\lambda}{m}\Big)
  \Bigg] \,, \\[10pt]
\begin{picture}(65,20)(1,1)
 \epsfxsize=2.cm \lower8pt \hbox{\epsfbox{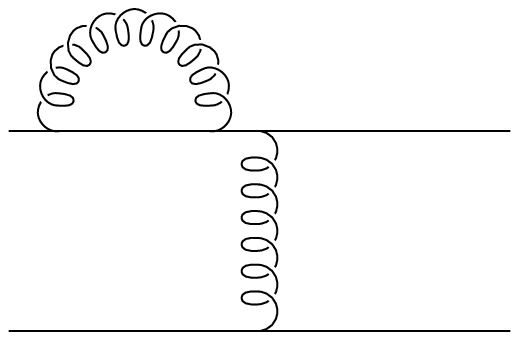}}
\end{picture} + \mbox{perms}\ \
 &=& \frac{i \alpha_s^2}{k^2}\, 2 C_F (T^A \otimes \bar T^A) \bigg\{
  2 \ln\Big(\frac{\lambda^2}{m^2}\Big) + \ln\Big( \frac{\mu^2}{m^2} \Big)
   +4 \bigg\} M^0  \,,\\[10pt]
\begin{picture}(65,20)(1,1)
 \epsfxsize=2.cm \lower14pt \hbox{\epsfbox{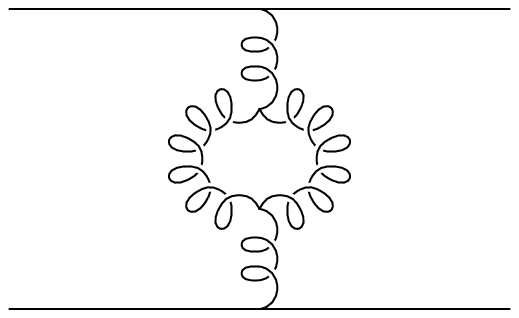}}
\end{picture}
+\
\begin{picture}(65,20)(1,1)
 \epsfxsize=2.cm \lower14pt \hbox{\epsfbox{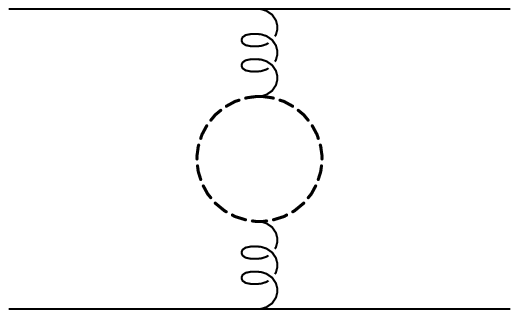}}
\end{picture}
 &=& -\frac{i \alpha_s^2}{k^2}\, C_A (T^A \otimes \bar T^A) \bigg\{ \frac53
 \ln\Big( \frac{\mu^2}{k^2} \Big)+\frac{31}{9} \bigg\} M^0 \,,\\[10pt]
\nonumber \\
 \noalign{\hbox{and the light quark loops for $n_f$ flavors gives:}}\nonumber
 \\*[0pt]
\begin{picture}(65,20)(1,1)
 \epsfxsize=2.cm \lower12pt \hbox{\epsfbox{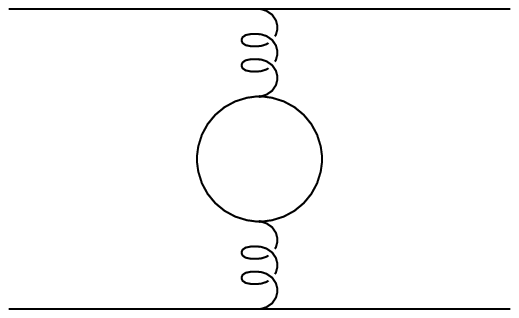}}
\end{picture}
 &=& \frac{i \alpha_s^2}{k^2}\, n_f T_F (T^A \otimes \bar T^A) \bigg\{
 \frac43 \ln\Big( \frac{\mu^2}{k^2} \Big) + \frac{20}{9} \bigg\} M^0 \,,\\[10pt]
 \nonumber \\[-10pt]
\noalign{\hbox{while the heavy quark fermion loop gives:}}\nonumber \\*
\begin{picture}(65,20)(1,1)
 \epsfxsize=2.cm \lower12pt \hbox{\epsfbox{qcd9.eps}}
\end{picture}
 &=& \frac{i \alpha_s^2}{k^2}\, T_F (T^A \otimes \bar T^A) \bigg\{
 \frac43 \ln\Big( \frac{\mu^2}{m^2} \Big) M^0 - \frac{4 k^2}{15 m^2} \bigg\}
 \,.
\end{eqnarray}
\end{mathletters}
In Eq.~(\ref{qcdother}) the matrix element
\begin{eqnarray}
  M^0 = 1+ \frac{p^2}{m^2} -\frac{k^2 {\bf S}^2}{3m^2}
  -\frac{3 k^2 \Lambda}{2 m^2}-\frac{k^2 {\bf T}}{12 m^2} \,.
\end{eqnarray}
Note that we disagree with Ref.~\cite{Yndurain} on the order $v^0$ spin
independent part of Eq.~(\ref{qcdother}b). Ref.~\cite{Yndurain} has an
additional non-logarithmic $1/m^2$ term. The difference arises because we find
a different value for the order $k^2$ term in the non-Abelian $F_1$
form-factor. Eqs.~(\ref{qcdother}c) through (\ref{qcdother}g) agree with
Ref.~\cite{Yndurain}.

The diagrams in Eq.~(\ref{box}) and (\ref{qcdother}) have contributions from
several different scales.  In particular, in the language of the threshold
expansion\cite{Beneke}, the hard regime gives $\ln(\mu/m)$'s,  the soft regime
gives $\ln(\mu/k)$'s, and the ultrasoft regime gives $\ln(\mu/\lambda)$'s.
There are also $i\ln(\lambda/p)$ and $i\ln(\lambda/k)$ terms from the Coulomb
divergence in the potential regime.  In addition to logarithms, all regimes can
give constant factors.  In the effective theory, terms from the hard regime are
absorbed into matching coefficients such as the four-quark potential operators
and the remaining terms correspond to  graphs involving modes in the effective
theory. These graphs are discussed in more detail below.

Next consider the one-loop annihilation diagrams in QCD. Since the intermediate
gluons are hard we expect these graphs to include a factor of $1/m^2$, thus
giving hard order $v^0$ contributions to the potential.  However, the graph in
Eq.~(\ref{annhil}c) also has terms enhanced by a factor of $m/p$, which are
order $1/v$ contributions.
\begin{mathletters} \label{annhil}
\begin{eqnarray}
 \begin{picture}(65,20)(1,1)
 \epsfxsize=2.cm \lower12pt \hbox{\epsfbox{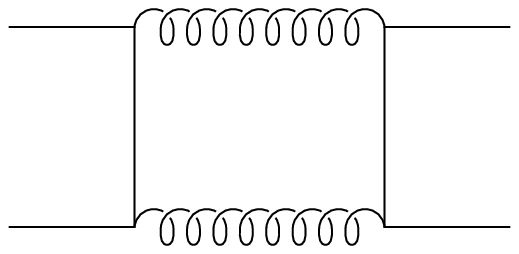}}
\end{picture}
+\
\begin{picture}(65,20)(1,1)
 \epsfxsize=2.cm \lower12pt \hbox{\epsfbox{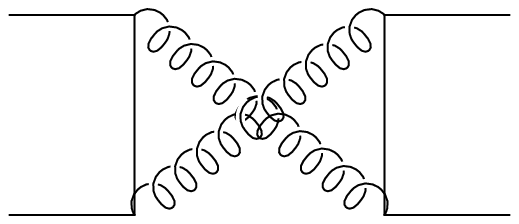}}
\end{picture}
&=& -\frac{i\alpha_s^2}{m^2}\Bigg[
 \bigg\{ \Big(\frac{C_1}{N_c}+\frac{C_d(N_c^2\!-\!1)}{8N_c^2}\Big)
  (1\otimes 1) +\Big( \frac{C_d}{4 N_c}\!-\!2 C_1\Big) (T^A\otimes \bar T^A)
  \bigg\} \nn\\*
  &\times& ( {\bf S^2}\!-\!2)(i\pi\!+\!2\!-\!2\ln 2)- C_A \bigg\{
   \frac{(N_c^2\!-\!1)}{2N_c^2} (1\otimes 1)\!+\!\frac{1}{N_c}(T^A \otimes
   \bar T^A) \bigg\} \nn\\*
  &\times& {\bf S^2} \bigg\{
  \frac{i\pi}{12}\!+\!\frac{1}{6}\!-\!
  \frac{\ln 2}{6}\!-\!\ln\Big(\frac{\lambda}{m}\Big) \bigg\} \Bigg]\,, \\
\begin{picture}(65,20)(1,1)
 \epsfxsize=2.cm \lower12pt \hbox{\epsfbox{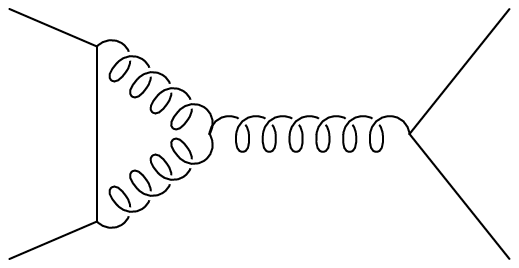}}
\end{picture}
+\
\begin{picture}(65,20)(1,1)
 \epsfxsize=2.cm \lower12pt \hbox{\epsfbox{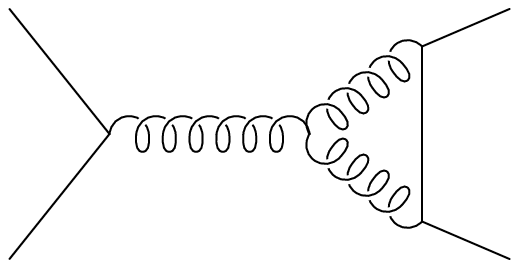}}
\end{picture}
&=& -\frac{i\alpha_s^2}{m^2}\: C_A\: \bigg\{ (T^A \otimes \bar T^A)
 \frac{1}{N_c} + ( 1\otimes 1) \frac{(N_c^2-1)}{2 N_c} \bigg\} \nn\\*
 &\times& {\bf S^2} \bigg\{ \frac{3}{2}\ln\Big(\frac{\mu}{m}\Big)
 + \frac{4}{3} +\frac{2\ln2}{3} -\frac{i\pi}{3} \bigg\} \,,  \\[5pt]
\begin{picture}(65,20)(1,1)
 \epsfxsize=2.cm \lower12pt \hbox{\epsfbox{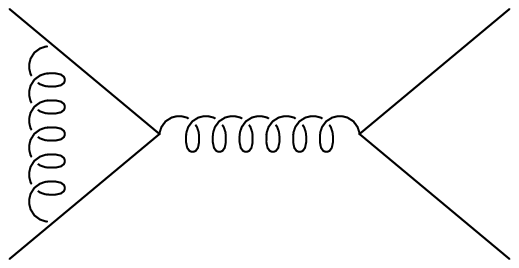}}
\end{picture}
+\
\begin{picture}(65,20)(1,1)
 \epsfxsize=2.cm \lower12pt \hbox{\epsfbox{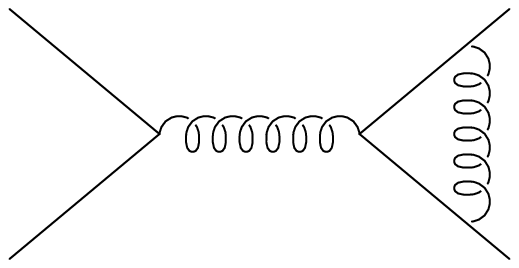}}
\end{picture}
 &=& -\frac{i \alpha_s^2 }{4m^2}\, (2C_F-C_A)\: \bigg\{ (T^A \otimes \bar T^A)
 \frac{1}{N_c} + ( 1\otimes 1) \frac{(N_c^2-1)}{2 N_c} \bigg\} \nn\\*
 &\times& {\bf S^2} \bigg\{
 \frac{\pi^2 m}{p} + \frac{2i\pi m}{p}\ln\Big(\frac{2p}{\lambda}\Big)
 - 4
 + 2 \ln\Big(\frac{\mu}{m}\Big) + 4 \ln\Big(\frac{\lambda}
 {m}\Big) \bigg\}  \,, \\
\begin{picture}(65,20)(1,1)
 \epsfxsize=2.cm \lower8pt \hbox{\epsfbox{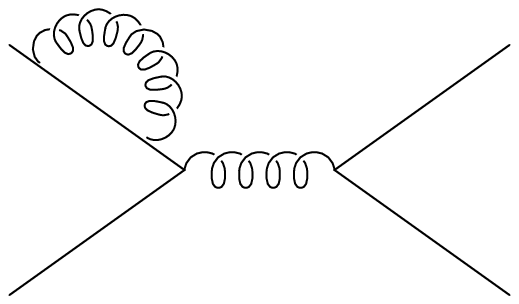}}
\end{picture} + \mbox{perms}\ \
 &=& \frac{i\alpha_s^2}{m^2}\: C_F \bigg\{ (T^A \otimes \bar T^A)
 \frac{1}{N_c} + ( 1\otimes 1)\frac{(N_c^2-1)}{2 N_c} \bigg\} \nn\\*
 &\times&  {\bf S^2}\: \bigg\{ \ln\Big(\frac{\lambda^2}
  {m^2}\Big) + \ln\Big( \frac{\mu}{m} \Big) +2  \bigg\} \,,\\
\begin{picture}(65,20)(1,1)
 \epsfxsize=2.cm \lower8pt \hbox{\epsfbox{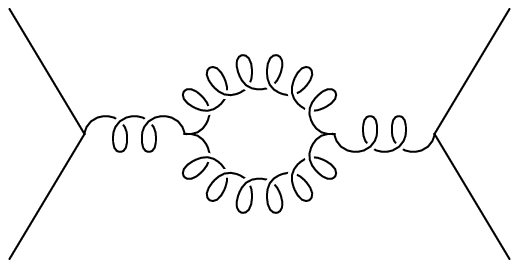}}
\end{picture} +
\begin{picture}(65,20)(1,1)
 \epsfxsize=2.cm \lower8pt \hbox{\epsfbox{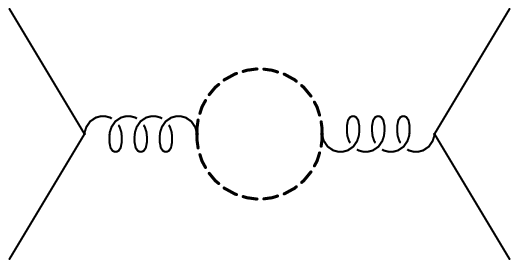}}
\end{picture}
 &=& -\frac{i\alpha_s^2}{m^2}\: \frac{5 C_A}{12} \bigg\{ (T^A \otimes\bar T^A)
 \frac{1}{N_c} + ( 1\otimes 1)\frac{(N_c^2-1)}{2 N_c} \bigg\} \nn\\*
 &\times&  {\bf S^2}\: \bigg\{ \ln\Big(\frac{\mu^2}{m^2}\Big)-2\ln 2
  +\frac{31}{15} + i \pi \bigg\} \,,\\
 \noalign{\hbox{and the light quark loop for $n_f$ flavors gives:}}
 \nonumber \\*
\begin{picture}(65,20)(1,1)
 \epsfxsize=2.cm \lower8pt \hbox{\epsfbox{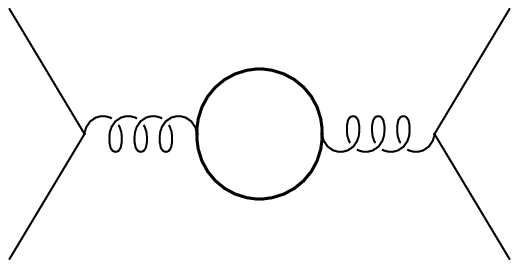}}
\end{picture}
 &=& \frac{i\alpha_s^2}{m^2}\: \frac{n_f T_F}{3} \bigg\{
 (T^A \otimes \bar T^A) \frac{1}{N_c} + ( 1\otimes 1)\frac{(N_c^2-1)}{2 N_c}
  \bigg\} \nn\\
 &\times&  {\bf S^2}\: \bigg\{ \ln\Big(\frac{\mu^2}{m^2}\Big) - 2\ln 2
 +\frac{5}{3} +i\pi \bigg\} \,, \\
\noalign{\hbox{while the heavy quark loop gives:}}\nonumber \\*
\begin{picture}(65,20)(1,1)
 \epsfxsize=2.cm \lower8pt \hbox{\epsfbox{qcda10.eps}}
\end{picture}
 &=& \frac{i\alpha_s^2}{m^2}\: \frac{T_F}{3} \bigg\{
 (T^A \otimes \bar T^A) \frac{1}{N_c} + ( 1\otimes 1)\frac{(N_c^2-1)}{2 N_c}
  \bigg\} \nn\\
 &\times&  {\bf S^2}\: \bigg\{ \ln\Big(\frac{\mu^2}{m^2}\Big) + \frac83
  \bigg\} \,.
\end{eqnarray}
\end{mathletters}
In the limit $p \to k/2$, the results in Eq.~(\ref{annhil}) agree with
Ref.~\cite{Gupta}, except for Fig.~(\ref{annhil}c) and differences that can be
accounted for by the fact that we are using \ms\ rather than an on-shell
subtraction scheme.  For Fig.~(\ref{annhil}c), Ref.~\cite{Gupta} has a $\pi^2/k$
term which for us is $-\pi^2/(2p)$. This sign for the $\pi^2/p$ term is in
agreement with Ref.~\cite{Andrep}. The imaginary parts of these one-loop
annihilation amplitudes also agree with Ref.~\cite{BBL}.  Note that in
Eqs.~(\ref{box}), (\ref{qcdother}), and (\ref{annhil}), $\alpha_s=\alpha_s(\mu)$.

\subsection{The Effective Theory Calculation}

The effective theory contains potential, soft and ultrasoft loops. We have
organized the terms by their order in the velocity expansion. The order in $v$
can be determined using the $v$ power counting formula\footnote{\tighten Here
the power of $v$ is given for the amputated diagram, so unlike Eq.~(40) in
Ref.~\cite{LMR} the factors of $v$ associated with external lines are not
included.} in Eq.~(40) of Ref.~\cite{LMR}. A loop graph with two insertions of
the Coulomb potential contributes to a $1/v^3$ potential, and with an insertion
of one Coulomb and one $V^{(0)}$ potential contributes to a $1/v$ potential. A
soft loop with two vertices of order $\sigma$ and $\sigma^\prime$ contributes
to the $v^{\sigma+\sigma^\prime-2}$ potential. Graphs involving the exchange of
an ultrasoft gluon begin to  contribute to the potential at order $v^0$, etc.

\subsubsection{Order $1/v^3$}

In the effective theory taking two insertions of the tree level Coulomb
potential in a loop gives the only diagram that is order $\alpha_s^2/v^3$:
\begin{eqnarray}
\begin{picture}(75,20)(1,1)
 \epsfxsize=2.4cm \lower9pt \hbox{\epsfbox{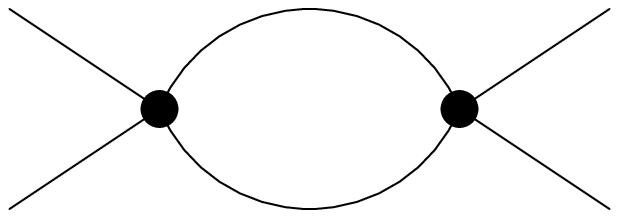}}
\end{picture}
  &=& i { [{\cal V}_c^{(T)}]^2 \over 16\pi^2 } (T^A T^B \otimes
  \bar T^A \bar T^B)\ { 2 i m \pi \over k^2\:p}  \ln\Big( \frac{{k}^2} {
  \lambda^2}\Big)   \,.
\end{eqnarray}
Taking $\mu=m$ and $\nu=1$ and using the tree level value of $V_c^{(T)}$, this
graph exactly reproduces the order $\alpha_s^2/v^3$ ``Coulomb singularity''
term in the QCD box diagram in Eq.~(\ref{box}). Thus, there is no matching
correction at this order.

\subsubsection{Order $1/v^2$}

At order $\alpha_s^2/v^2$ in the effective theory, the only non-zero diagram
involves the exchange of soft gluons, ghosts and quarks:
\begin{eqnarray}  \label{eftvm2}
\begin{picture}(60,30)(1,1)
 \epsfxsize=2.0cm \lower16pt \hbox{\epsfbox{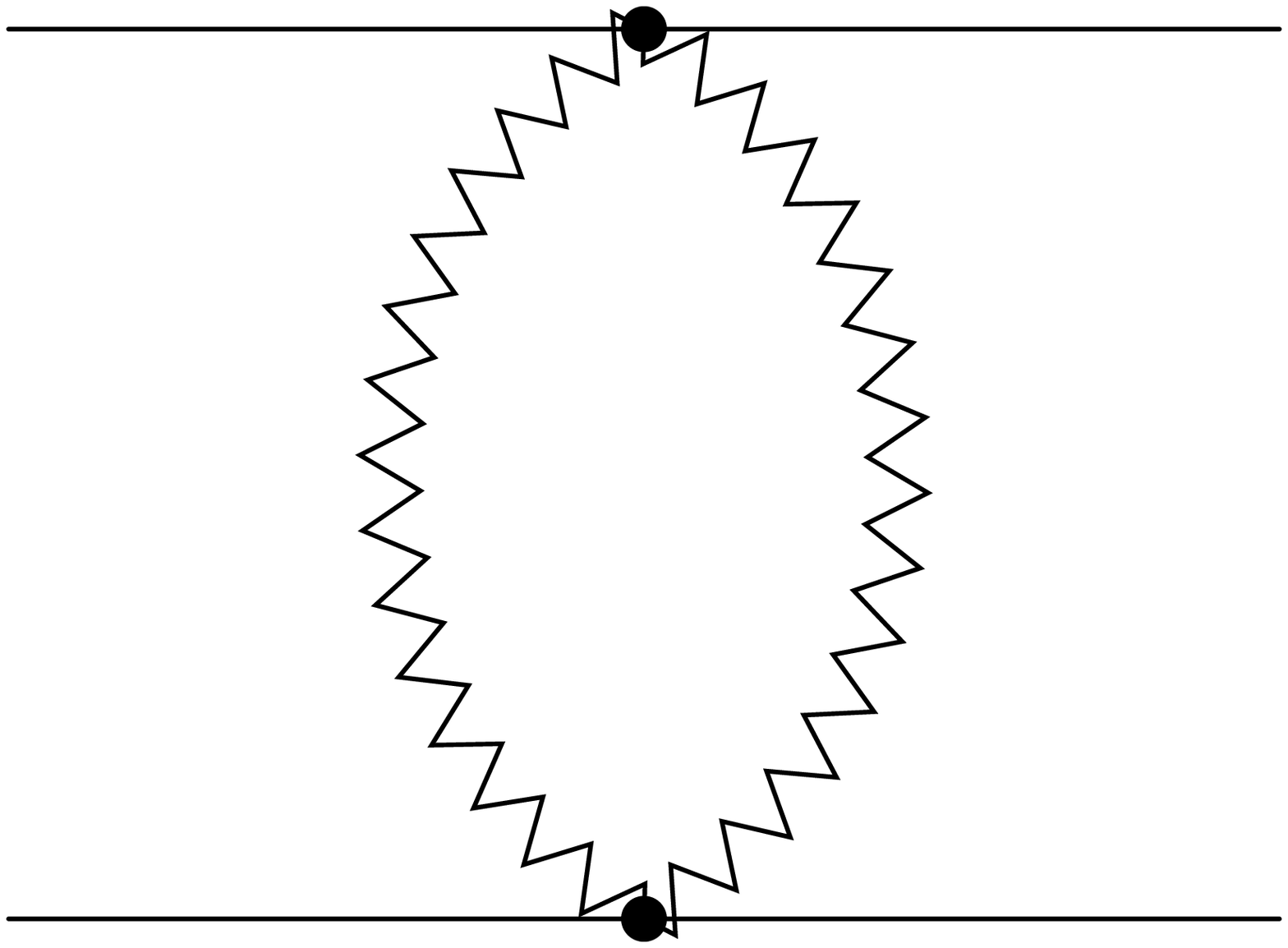}}
\end{picture}
  &=& \frac{i \alpha_s^2(\mu_S)}{k^2} (T^A  \otimes \bar T^A) \Bigg[
   \frac{4 T_F n_f-11 C_A}{3} \bigg\{ \frac{1}{\hat\epsilon}+  \ln\Big(
   \frac{\mu_S^2}{k^2}\Big) \bigg\} +\frac{20 T_F n_f -31 C_A}{9} \Bigg] \,,
\end{eqnarray}
where $1/\hat\epsilon=1/\epsilon-\gamma+\ln(4\pi)$ is the combination
subtracted in \ms. The divergence in this graph is responsible for the one-loop
running of ${\cal V}_c^{(T)}$. At one-loop, to order $v^0$ the effective theory
counterterms (and running of the potential) were computed in Ref.~\cite{amis}
and from now on the $1/\hat\epsilon$ dependence will be dropped. The sum of
order $\alpha_s^2/v^2$ terms from the QCD diagrams in Eqs.~(\ref{box}) and
(\ref{qcdother}) is
\begin{eqnarray}  \label{qcdvm2}
  \frac{i \alpha_s^2(\mu)}{k^2} (T^A  \otimes \bar T^A) \Bigg[
   \frac{4 T_F}{3}  \ln\Big(\frac{\mu^2}{m^2}\Big) +
   \frac{4 T_F n_f-11 C_A}{3} \ln\Big(
   \frac{\mu^2}{k^2}\Big) +\frac{20 T_F n_f -31 C_A}{9} \Bigg]    \,.
\end{eqnarray}
At the scale $\mu=\mu_S= m$ the \ms\ values of Eq.~(\ref{eftvm2}) and
Eq.~(\ref{qcdvm2}) are identical, so there is no one-loop matching correction
to ${\cal V}_c$. The difference in $\ln(\mu)$'s in Eqs.~(\ref{qcdvm2}) and
(\ref{eftvm2}) is the change in $\beta$-function from $n_f+1$ to $n_f$ flavors.
We would expect a new contribution to ${\cal V}_c$ only if the QCD graphs have
a contribution from the hard regime where energy and momenta are order $m$. In
Eqs.~(\ref{qcdother}b), (\ref{qcdother}c), and (\ref{qcdother}d), the factors
of $-3\ln(m^2)+4$ are from the hard regime, but they sum to zero.  For the
$\ln(m)$ terms this is a consequence of the Ward identity derived in
Ref.~\cite{adm}.  The constant factor $(20 T_F n_f-31 C_A)/9$ vanishes in the
matching condition at $m$, and is carried to scales below $m$ by the soft
modes.

Note that if at the scale $\nu=v_k\sim k/m$ we integrate out the soft modes,
then the Coulomb potential for the theory below this scale, $\overline {\cal
V}_c$, will obtain an additional contribution from this second stage of
matching:  $\overline {\cal V}_c^{(T)}(v_k) = {\cal V}^{(T)}_c(v_k) - (20 T_F
n_f-31 C_A)\alpha_s^2(m v_k)/9$. In this expression ${\cal V}^{(T)}_c(v_k)$ is
the value of ${\cal V}^{(T)}_c$ obtained from running this coefficient from
$\nu=1$ to $\nu=v_k$ using its two-loop anomalous dimension. This reproduces
the constant factor that is typically associated with the Coulomb potential at
next-to-leading order (see for instance, Ref.~\cite{Peter}). Similarly, one can
obtain the additional terms from the matching contributions for the $v^0$
potentials at $mv$ from the value of the soft loops given in
Eq.~(\ref{eftv0soft}).

\subsubsection{Order $1/v$}

The possible order $\alpha_s^2/v$ diagrams with soft gluons vanish explicitly,
so the only order $\alpha_s^2/v$ diagrams in the effective theory involve two
iterations of the potential.  There are two diagrams with insertions of ${\cal
V}_c^{(T)}$:
\begin{eqnarray} \label{eft1}
 \begin{picture}(75,20)(-5,1)
   \put(9,16){${\cal V}_c$} \put(48,16){${\cal V}_c$}
   \put(30,11.5){${\large \mathbf{\times}}$}
   \epsfxsize=2.4cm \lower9pt \hbox{\epsfbox{eft1.eps}}
 \end{picture} +\ \Delta E
 \begin{picture}(75,20)(-5,1)
   \put(9,16){${\cal V}_c$} \put(48,16){${\cal V}_c$}
   \epsfxsize=2.4cm \lower9pt \hbox{\epsfbox{eft1.eps}}
 \end{picture}
   &=& \frac{i [{\cal V}_c^{(T)}]^2 }{4 m}\: T^A T^B \otimes \bar T^A \bar T^B
    \, \Big( I_F + 2\, {p}^2\, I_0 \Big)
 \,,
\end{eqnarray}
where the integrals $I_0$ and $I_F$ are given in Appendix~\ref{App_int}. The
dependence on ${\cal V}_c^{(1)}$ is not needed since tree level matching
gives ${\cal V}_c^{(1)}=0$.  In the first diagram in Eq.~(\ref{eft1}), the cross
denotes an insertion of the ${\bf p}^4/m^3$ operator from Eq.~(\ref{Lu}). The
second diagram is nominally order $\alpha_s^2/v^3$; however it depends on the
heavy quark energy $E={\bf p}^2/(2m)-{\bf p}^4/(8 m^3)+ \ldots$. When the energy
is expanded in terms of momenta, the graph includes a contribution of order
$\alpha_s^2/v$ which we indicate by the  pre-factor $\Delta E$ in
Eq.~(\ref{eft1}). Each of the diagrams in Eq.~(\ref{eft1}) has an IR divergence
that is not regulated by $\lambda$, but the IR divergence in the sum of the
integrands for the two diagrams is regulated.

Additional $\alpha_s^2/v$ diagrams are generated by including one insertion of
the Coulomb potential and one order $v^0$ potential:
\begin{mathletters} \label{eftVcV0}
\begin{eqnarray}  
 \begin{picture}(75,40)(-5,1)
   \put(9,16){${\cal V}_c$} \put(48,16){${\cal V}_r$}
   \put(50,-1.7){$\Box$}
   \epsfxsize=2.4cm \lower9pt \hbox{\epsfbox{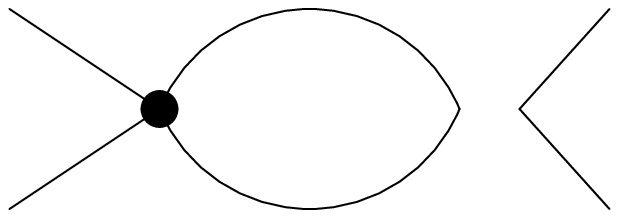}}
 \end{picture} +
 \begin{picture}(75,40)(-5,1)
   \put(9,16){${\cal V}_r$} \put(48,16){${\cal V}_c$}
   \put(9,-1.7){$\Box$}
   \epsfxsize=2.4cm \lower9pt \hbox{\epsfbox{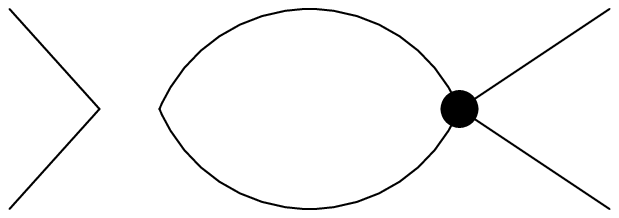}}
 \end{picture}
   &=& \frac{i {\cal V}_c^{(T)} {\cal V}_r^{(T)} }{m}\: T^A T^B \otimes
      \bar T^A \bar T^B\, \Big( I_F + 2\, {p}^2\, I_0 \Big)
 \,, \\
 \begin{picture}(75,40)(-5,1)
   \put(9,16){${\cal V}_c$} \put(48,16){${\cal V}_s$}
   \put(50,-1.7){$\Box$}
   \epsfxsize=2.4cm \lower9pt \hbox{\epsfbox{eft2.eps}}
 \end{picture} +
 \begin{picture}(75,40)(-5,1)
   \put(9,16){${\cal V}_s$} \put(48,16){${\cal V}_c$}
   \put(9,-1.7){$\Box$}
   \epsfxsize=2.4cm \lower9pt \hbox{\epsfbox{eft3.eps}}
 \end{picture}
   &=& \frac{i {\cal V}_c^{(T)} {\cal V}_s^{(T)} }{m}\: T^A T^B \otimes
    \bar T^A \bar T^B\: \Big( 2 I_P \:  {\mathbf S}^2 \Big)
 \,, \\
 \begin{picture}(75,40)(-5,1)
   \put(9,16){${\cal V}_c$} \put(48,16){${\cal V}_{\Lambda}$}
   \put(50,-1.7){$\Box$}
   \epsfxsize=2.4cm \lower9pt \hbox{\epsfbox{eft2.eps}}
 \end{picture} +
 \begin{picture}(75,40)(-5,1)
   \put(9,16){${\cal V}_{\Lambda}$} \put(48,16){${\cal V}_c$}
   \put(9,-1.7){$\Box$}
   \epsfxsize=2.4cm \lower9pt \hbox{\epsfbox{eft3.eps}}
 \end{picture}
   &=& \frac{i {\cal V}_c^{(T)} {\cal V}_{\Lambda}^{(T)} }{m}\: T^A T^B \otimes
     \bar T^A \bar T^B\, \Big( 2 {k}^2\, \Lambda({\bf p',p})\: I_A
     \Big)
 \,, \\
 \begin{picture}(75,40)(-5,1)
   \put(9,16){${\cal V}_c$} \put(48,16){${\cal V}_t$}
   \put(50,-1.7){$\Box$}
   \epsfxsize=2.4cm \lower9pt \hbox{\epsfbox{eft2.eps}}
 \end{picture} +
 \begin{picture}(75,40)(-5,1)
   \put(9,16){${\cal V}_t$} \put(48,16){${\cal V}_c$}
   \put(9,-1.7){$\Box$}
   \epsfxsize=2.4cm \lower9pt \hbox{\epsfbox{eft3.eps}}
 \end{picture}
   &=& \frac{-i {\cal V}_c^{(T)} {\cal V}_t^{(T)} }{2m}\: T^A T^B \otimes
    \bar T^A \bar T^B\, \bigg[{\mathbf k\cdot\sigma_1\, k\cdot\sigma_2}
  (12 I_C+ 3 I_0) \nn\\*
   &-4& {\mathbf \sigma_1\cdot\sigma_2}
    (I_P-3 I_B) + {\mathbf R}
  (12 I_D-12 I_A + 3 I_0) \bigg]
 \,, \\[-5pt]
 \begin{picture}(75,40)(-5,1)
   \put(9,16){${\cal V}_c$} \put(42,16){${\cal V}_{s,a}$}
   \put(50,-1.7){$\Box$}
   \epsfxsize=2.4cm \lower9pt \hbox{\epsfbox{eft2.eps}}
 \end{picture} +
 \begin{picture}(75,40)(-5,1)
   \put(3,16){${\cal V}_{s,a}$} \put(48,16){${\cal V}_c$}
   \put(9,-1.7){$\Box$}
   \epsfxsize=2.4cm \lower9pt \hbox{\epsfbox{eft3.eps}}
 \end{picture}
   &=& \frac{i {\cal V}_c^{(T)} {\cal V}_{s,a}^{(T)} }{m}\: T^A T^B \otimes
    \bar T^A \bar T^B\ \Big( {\bf S^2} \ 2 I_P \Big) \nn\\*
   && + \frac{i {\cal V}_c^{(T)} {\cal V}_{s,a}^{(1)} }{m}\: T^A \otimes
    \bar T^A \ \Big( {\bf S^2} \ 2 I_P \Big) \,.
\end{eqnarray}
\end{mathletters}
The last two diagrams involve insertions of terms in the potential generated by
the tree level annihilation diagram. In Eqs.~(\ref{eft1}) and (\ref{eftVcV0})
the dependence on $1\otimes 1$ potentials whose coefficients
vanish at tree level are not shown. Thus, both the $1\otimes 1$ and $T\otimes
\bar T$ contributions are only shown in Eq.~(\ref{eftVcV0}e). The new integrals
$I_P$, $I_A$, $I_B$, $I_C$, and $I_D$ that appear in Eq.~(\ref{eftVcV0}) are
given in Appendix~\ref{App_int}, and $\mathbf R$ is defined in
Eq.~(\ref{Rdef}).

The infrared divergences in the $1/v$ amplitude are due to the Coulomb
singularity. The divergences in the full theory are reproduced by the potential
loops in Eqs.~(\ref{eft1}) and (\ref{eftVcV0}). The imaginary terms in the
amplitude from the QCD box graph exactly agree with those in the effective
theory, and do not contribute in the matching coefficients, as expected.
Subtracting the effective theory graphs in Eqs.~(\ref{eft1}) and (\ref{eftVcV0})
from the order $\alpha_s^2/v$ terms in Eq.~(\ref{box}), (\ref{qcdother})a,b and
(\ref{annhil})c gives the matching result for the $V^{(-1)}$ potential at
$\mu=m$, $\nu=1$:
\begin{eqnarray} \label{Lkresult}
  {\cal V}_k^{(T)} &=& \alpha_s^2(m) \Big( \frac{7 C_A}{8}-\frac{C_d}{8}
  \Big) \,,\qquad
   {\cal V}_k^{(1)} = \alpha_s^2(m) \frac{C_1}{2} \,.
\end{eqnarray}
For the color singlet channel this gives the matching coefficient ${\cal
V}_k^{(s)}=\alpha_s^2(m) (C_F^2/2-C_F C_A)$ in agreement with
Ref.~\cite{Yndurain}.  Note that the real part of the $1/v$ amplitudes in the
full and effective theory have a complicated dependence on $p$ and $k$, but the
momentum dependence of the matching for the direct potentials in
Eq.~(\ref{Lkresult}) is only of the form $1/\abs{\mathbf k}$.
For the annihilation graphs the $1/p$ terms cancel between the full and
effective theories. There would have been a $1/p$ matching coefficient for the
annihilation graphs if the results of Ref.~\cite{Gupta} had been used for the
full theory graph.

The matching coefficients in Eq.~(\ref{Lkresult}) correspond to a contribution
to the $1/v$ potential at the scale $\mu=m$, which does not come from the hard
part of any graph. This is in apparent contradiction with the threshold
expansion.  It also appears to disagree with Refs.~\cite{Brambilla} and
\cite{BSS}, where the $1/|{\bf k}|$ four-quark potential operator is said to
only arise at the scale $mv$. However, in Refs.~\cite{Brambilla} and \cite{BSS}
a formulation of NRQCD is being used in which off-shell potential field
components have not been integrated out for scales $m v < \mu < m$, but instead
are considered to be dynamical fields in the Lagrangain (see
Refs.~\cite{lm,ls}). In the full theory the $1/|{\bf k}|$ terms come from three
types of graphs, shown in Eq.~(\ref{box}) and Eqs.~(\ref{qcdother}a,b).  At the
scale $m$ there are now effective theory graphs analogous to those in
Eq.~(\ref{qcdother}b) but with two $A^0$ potential gluons and one ${\bf A}^i$
potential gluon, which reproduce the $1/|{\bf k}|$ term.  For the box and
crossed box in Feynman gauge the $1/|{\bf k}|$ terms are reproduced by
contributions that can be associated with the potential momentum regime. Thus,
in Refs.~\cite{Brambilla} and \cite{BSS} the $1/|{\bf k}|$
potential in Eq.~(\ref{Lkresult}) effectively exists at the scale $m$. In our
approach off-shell potential gluons and soft quarks are integrated out, so the
matching for the  $1/|{\bf k}|$ potential is not simply given by the hard part
of the QCD diagrams.

\subsubsection{Order $v^0$}

For the order $\alpha_s^2 v^0$ matching we will consider the direct and
annihilation diagrams separately.  The sum of the order $\alpha_s^2 v^0$ terms
in the direct scattering QCD diagrams in Eqs.~(\ref{box}) and  (\ref{qcdother})
is
\begin{eqnarray} \label{qcdv0sum}
 -\frac{i\alpha_s^2(\mu)}{m^2} \ \ && (1\otimes 1) C_1 \Bigg[ 4 -
   2 {\bf S^2} + 2 \ln\Big(\frac{k}{m}\Big)
    -\frac{8}{3} \ln\Big(\frac{\lambda}{k}\Big) \Bigg] \\
 -\frac{i\alpha_s^2(\mu)}{m^2}\ \ && (T^A \otimes\bar T^A) \Bigg[
 \frac{{p}^2}{k^2} \Bigg\{
 \frac{31C_A}{9}+\frac{16C_A}{3} \ln\Big(\frac{\lambda}{\mu}\Big) +
 \frac{38C_A}{3}\ln\Big(\frac{\mu}{k}\Big) \Bigg\} \nn\\
  && \qquad\qquad\quad + \Bigg\{ 2 C_F -3 C_A-C_d +
  \Big(\frac{8C_F}{3}+\frac{2 C_d}{3}-2 C_A \Big)
  \ln\Big(\frac{\lambda}{\mu}\Big) \nn\\
  &&\qquad\qquad\quad + \Big( \frac{7 C_d}{6}-\frac{43
  C_A}{6}\Big) \ln\Big(\frac{\mu}{k}\Big) + \Big( \frac{8 C_F}{3}+\frac{31
  C_A}{6}-\frac{C_d}{2} \Big) \ln\Big( \frac{\mu}{m} \Big) \Bigg\} \nn\\
 && \qquad\qquad\quad + \Lambda \Bigg\{ -\frac{31\, C_A}{6}-4 C_F
 -7 C_A \ln\Big(\frac{\mu}{k}\Big) -4 C_A \ln\Big(\frac{\mu}{m} \Big) \Bigg\}
  \nn\\
 && \qquad\qquad\quad + {\bf S^2} \Bigg\{ \frac{C_d}{2} -\frac{17
 C_A}{54}-\frac{4 C_F}{3} -\frac{C_A}{9} \ln\Big(\frac{\mu}{k}\Big)
 -\frac{7C_A}{3} \ln\Big(\frac{\mu}{m}\Big) \Bigg\} \nn\\
 && \qquad\qquad\quad + {\bf T} \Bigg\{ -\frac{49 C_A}{108} -
 \frac{C_F}{3} -\frac{5 C_A}{18} \ln\Big(\frac{\mu}{k}\Big)
 -\frac{C_A}{3} \ln\Big(\frac{\mu}{m}\Big) \Bigg\} \nn\\
 &&\qquad\qquad\quad - n_f T_F \bigg\{\frac43 \ln\Big( \frac{\mu^2}{k^2}\Big)
  + \frac{20}{9} \bigg\} \bigg\{ \frac{p^2}{k^2} -
  \frac{ {\bf S}^2}{3}-\frac{3  \Lambda}{2 }-
  \frac{ {\bf T}}{12 }\bigg\} \nn\\
 &&\qquad\qquad\quad - \frac{4T_F}{3} \ln\Big( \frac{\mu^2}{m^2}\Big)
  \bigg\{ \frac{ p^2}{k^2} -
  \frac{{\bf S}^2}{3}-\frac{3  \Lambda}{2 }-
  \frac{{\bf T}}{12 }\bigg\} + \frac{4 T_F }{15 } \Bigg] \nn \,.
\end{eqnarray}
The effective theory contribution from order $\alpha_s^2 v^0$ diagrams with
soft gluons, ghosts or quarks is:
\begin{eqnarray}  \label{eftv0soft}
\begin{picture}(60,30)(1,1)
 \epsfxsize=2.0cm \lower16pt \hbox{\epsfbox{eft_soft.eps}}
\end{picture}
 &=& \frac{i \alpha_s^2(\mu_S)}{m^2}\ ({1\otimes 1})\Bigg[ { {14 C_1\over 3}}\,
  \ln\Big(\frac{\mu_S}{k}\Big)-\frac{C_1}{3} \Bigg] \\*
  &+& \frac{i \alpha_s^2(\mu_S)}{m^2}\ (T^A  \otimes \bar T^A) \Bigg[
   C_A \ln\Big(\frac{\mu_S}{k}\Big) \Bigg\{ {43 \over 6 } -
   {38 p^2  \over 3 k^2 }
   +  {7\Lambda} + { {\mathbf S}^2 \over 9 }  +{5 {\bf T} \over 18 }
   \Bigg\}\nn\\*
  && -{7 C_d \over 6 }\ln\Big(\frac{\mu_S}{k}\Big)+\frac{C_d}{12}
  +\frac{7 C_A}{3}-\frac{31 C_A p^2}{9 k^2}+
  \frac{7 C_A \Lambda}{6}-\frac{14 C_A {\bf S^2}}{27} +\frac{13 C_A {\bf T}}
  {108}\nn \\*[10pt]
  && + n_f T_F \bigg\{\frac43 \ln\Big( \frac{\mu_S^2}{k^2}\Big)
  + \frac{20}{9} \bigg\} \bigg\{ \frac{p^2}{ k^2} -
  \frac{{\bf S}^2}{3}-\frac{3 \Lambda}{2 }-
  \frac{ {\bf T} }{12}\bigg\} \Bigg]\nn \,.
\end{eqnarray}
There are also non-zero order $\alpha_s^2 v^0$ diagrams with an ultrasoft gluon
and one insertion of the Coulomb potential. These diagrams were calculated in
Ref.~\cite{amis}:
\begin{eqnarray} \label{eftv0usoft}
\begin{picture}(60,30)(1,1)
 \epsfxsize=2.0cm \lower16pt \hbox{\epsfbox{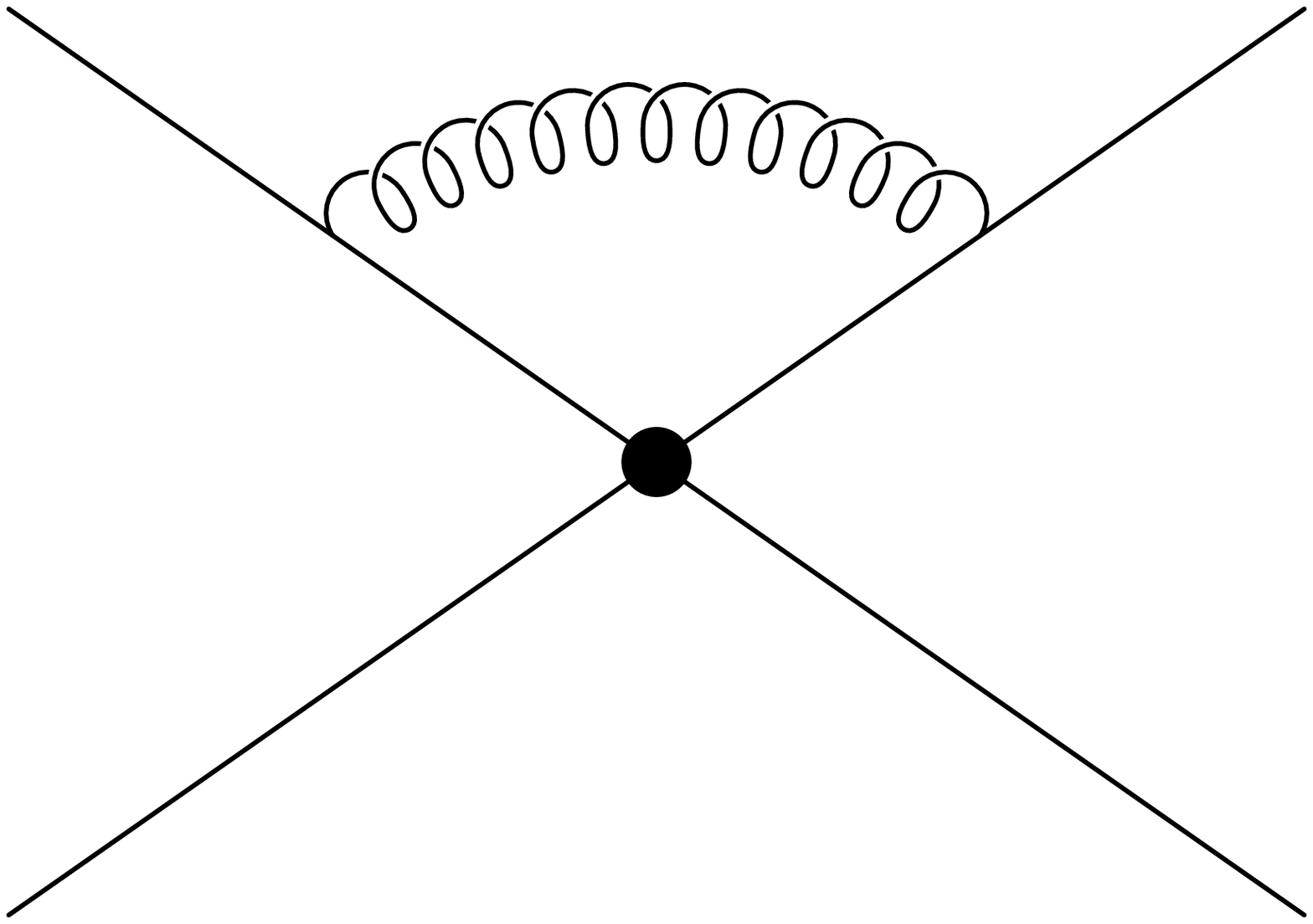}}
\end{picture} +\quad \ldots\quad
&=& \frac{8i}{3m^2}\: \alpha_s(\mu_S)\alpha_s(\mu_U)\:
  \ln\Big(\frac{\mu_U}{\lambda}\Big) \Bigg[ -C_1 (1\otimes 1)\nn\\*
  &&  + (T\otimes \bar T)\Bigg\{  C_F +\frac{C_d}{4}
  -\frac{3 C_A}{4} + \frac{2p^2 C_A}{ k^2}  \Bigg\} \Bigg]
  \,.
\end{eqnarray}
In Eq.~(\ref{eftv0usoft}) the ellipses denote all possible  diagrams that are
generated by attaching the ultrasoft gluon to two fermion  legs.  Subtracting
the sum of Eq.~(\ref{eftv0soft}) and (\ref{eftv0usoft})  from
Eq.~(\ref{qcdv0sum}) and setting $\mu=\mu_S=\mu_U=m$ gives the  one-loop
matching contribution for the order $v^0$ direct potentials. At order $v^0$ we
find that all the infrared divergences in QCD are matched by infrared
divergences from the effective theory graphs with ultrasoft gluons. All
$\ln(k)$'s in the full theory are matched by $\ln(k)$'s in
Eq.~(\ref{eftv0soft}). The contributions to the matching coefficients at one-loop
are:
\begin{eqnarray} \label{fmatch} {\normalsize
 \begin{tabular}{lll}
  ${\cal V}_r^{(T)}  = 0$ \,, & \phantom{ccccc} &
  ${\cal V}_r^{(1)}  = 0$ \,, \\
  ${\cal V}_\Lambda^{(T)} = -4(C_F+C_A)\, \alpha_s^2(m)$\,, & &
  ${\cal V}_\Lambda^{(1)} = 0$ \,, \\
  ${\cal V}_s^{(T)} = \Big( \frac{1}{2}C_d-\frac{5}{6}C_A-\frac{4}{3}C_F \Big)
   \,\alpha_s^2(m)$ \,, & &
  ${\cal V}_s^{(1)} = -2 C_1\: \alpha_s^2(m)$ \,, \\
  ${\cal V}_t^{(T)} =  -\frac{1}{3}(C_F+C_A)\, \alpha_s^2(m) $ \,, & &
  ${\cal V}_t^{(1)} = 0 $ \,, \\
  ${\cal V}_2^{(T)} =  \Big( 2 C_F - \frac{11}{12} C_d-\frac{2}{3} C_A +
   \frac{4}{15} T_F \Big)\, \alpha_s^2(m)$ \,, & &
  ${\cal V}_2^{(1)} = \frac{11}{3}\, C_1\: \alpha_s^2(m) $ \,.
 \end{tabular}  }
\end{eqnarray}
The total order $\alpha_s^2 v^0$ matching coefficients are given by adding
the tree level values in Eq.~(\ref{bc}) to the results in Eq.~(\ref{fmatch}),
and are summarized in Table~\ref{tab:results} at the end of the paper.

Next consider the matching of the order $\alpha_s^2 v^0$ annihilation
contributions. Since there are no corresponding diagrams in the effective
theory, the sum of the $\alpha_s^2 v^0$ terms in Eq.~(\ref{annhil}) directly
give the matching coefficients. This sum is infrared finite.  Matching at
$\mu=m$, $\nu=1$ we find that the one-loop annihilation contributions to the
potential coefficients are:
\begin{eqnarray} \label{amatch}
 {\cal V}_{s,a}^{(T)} &=&  \Big( \frac{C_d}
   {4N_c} -2 C_1 \Big) (i\pi+2-2\ln 2)\: \alpha_s^2(m) \nn\\*
   && +  \frac{1}{N_c} \bigg\{ \frac{109 C_A}{36}-4 C_F  + \frac{n_f T_F}{3}
   \Big( 2\ln 2 -\frac{5}{3} -i\pi \Big) -\frac{8 T_F}{9} \bigg\}\:
   \alpha_s^2(m) \,, \nn\\*[5pt]
 {\cal V}_{s,a}^{(1)} &=&
   \bigg\{ \frac{C_1}{N_c} +\frac{C_d(N_c^2-1)}{8 N_c^2} \bigg\}
   (i\pi+2-2\ln 2)\: \alpha_s^2(m) \nn\\*
    && +  \frac{(N_c^2-1)}{2N_c^2} \bigg\{ \frac{109 C_A}{36}-4 C_F  +
   \frac{n_f T_F}{3}\Big( 2\ln 2 -\frac{5}{3} -i\pi \Big) -\frac{8 T_F}{9}
   \bigg\}\: \alpha_s^2(m)  \,,\nn\\[5pt]
 {\cal V}_{2,a}^{(T)} &=& -2 \Big( \frac{C_d}{4N_c} -2 C_1 \Big)
   (i\pi+2-2\ln 2)\: \alpha_s^2(m)  \,,\nn\\*[5pt]
 {\cal V}_{2,a}^{(1)} &=& -2 \bigg\{ \frac{C_1}{N_c} +\frac{C_d(N_c^2-1)}
   {8 N_c^2} \bigg\} (i\pi+2-2\ln 2)\: \alpha_s^2(m) \,.
\end{eqnarray}
The imaginary terms in these potentials contribute to the cross section for
annihilation of a color octet heavy quark and anti-quark into light hadrons, and
agree with the results of Ref.~\cite{BBL}. The total annihilation contribution
to the order $\alpha_s^2 v^0$ matching coefficients are given by adding
the tree level results in Eq.~(\ref{bc2}) to the results in Eq.~(\ref{amatch}),
and are summarized in Table~\ref{tab:results}.

If a different matching scale, $\mu_h$, had been used then the coefficients in
Eqs.~(\ref{fmatch}) and (\ref{amatch}) would also depend on $\ln(\mu_h/m)$.
Since the prediction for observables is independent of $\mu_h$ the most
convenient choice, $\mu_h=m$, has been adopted.

In the threshold expansion, the full QCD diagram is the sum of hard, soft,
ultrasoft and potential graphs. The soft, ultrasoft and potential graphs are the
graphs in the effective theory up to renormalization effects such as the
dependence on the scales $\mu_S$ and $\mu_U$. The matching conditions in
Eq.~(\ref{fmatch}) and (\ref{amatch}) can also be computed from the hard part of
the full theory graphs, and we have verified that this gives the same result as
in Eqs.~(\ref{fmatch}) and (\ref{amatch}). The matching onto four-quark
operators was first considered by Pineda and Soto in the context of
pNRQCD~\cite{P2}. In their approach, the hard parts of the vertex correction and
wavefunction renormalization are included in matching coefficients in the single
heavy quark sector of the Lagrangian. The direct matching for four-quark
operators is given by the hard part of the box and crossed-box.  The hard part
of our box and crossed-box agree with Ref.~\cite{P2}, except for the finite part
of the ${\mathbf \sigma_1 \cdot \sigma_2}$ terms, which causes our ${\cal
V}_2^{(T)}$ and ${\cal V}_s^{(T)}$ to disagree with theirs. This is
related to the treatment of epsilon tensors and the non-relativistic reduction
of matrix elements of spin operators in $d$ dimensions. We have chosen to take
the lowest order term in the matrix element of $\gamma^{[\alpha} \gamma^\sigma
\gamma^{\beta]} \otimes \gamma_{[\alpha} \gamma_\tau \gamma_{\beta]}$ to be
$\epsilon^{\alpha\beta\sigma k} \epsilon_{\alpha\beta\tau k'}\ {\mathbf
\sigma}_1^k {\mathbf \sigma}_2^{k'}$, and have used the 't~Hooft-Veltmann
scheme for the epsilon tensors so that $\epsilon_{\mu\nu\alpha\beta}$ and
$\epsilon^{ijk}$ are only non-zero when the indices are in four and three
dimensions respectively. This convention was used in both the soft loop
integrals as well as when cross checking our result by computing the hard part
of the box diagrams using the threshold expansion.  This scheme dependence is
related to the issue of evanescent operators \cite{Buras,Dugan,Herrlich}. The
annihilation results in Eq.~(\ref{amatch}) agree completely with Ref.~\cite{P2}.

Note that in a leading-log expansion the two-loop anomalous dimension is needed
at the same time as the one-loop matching results.  In the color-singlet
channel the $1/v^2$ two-loop anomalous dimensions is
known~\cite{Peter,Schroder}, since the running of the Coulomb potential is
still given by the QCD $\beta$-function at this order. The result for the
two-loop anomalous dimension for $V^{(-1)}$ will be presented in another
publication\cite{amis3}.

\section{Terms in the potential which vanish on-shell.}

For the scattering  $Q(p^0_1,{\bf p})+{\bar Q}(p^0_2,-{\bf p})\to Q(p^0_3,{\bf
p^\prime}) + {\bar Q}(p_4^0,-{\bf p^\prime})$ consider adding
\begin{eqnarray} \label{V0a}
  V^{(0)} &=&
  \bigg[ {\cal V}_{\Delta 1}^{(T)}\: (T^A \otimes \bar T^A) +
  {\cal V}_{\Delta 1}^{(1)}\: (1\otimes 1) \bigg]
  { (p_3^0-p_1^0)^2 \over {\mathbf k}^4 }\ \nn\\[5pt]
  &+& \bigg[ {\cal V}_{\Delta 2}^{(T)}\: (T^A \otimes \bar T^A) +
  {\cal V}_{\Delta 2}^{(1)}\: (1\otimes 1) \bigg]
  { {(\mathbf p^{\prime 2} - p^2})^2 \over 4 m^2 {\mathbf k}^4 }\
\end{eqnarray}
to the potential in Eq.~(\ref{V0}). Here ${\bf k}={\bf p^\prime}-{\bf p}$ and
by energy conservation $p_3^0-p_1^0=p_2^0-p_4^0$. On-shell the potentials in
Eq.~(\ref{V0a}) vanish, since $p_1^0=p_3^0$, $p_4^0=p_2^0$, and ${\bf p^2}={\bf
p'}^2$. However, if we work off-shell, they are valid terms and in fact show up
in many calculations that make use of time-ordered perturbation theory.

Matching to the tree level diagrams in Fig.~\ref{fig_tree} with $p_1^0\ne
p_3^{0}$ gives a contribution to the potentials in Eq.~(\ref{V0a}). In Feynman
gauge one finds
\begin{eqnarray} \label{VdF}
  {\cal V}_{\Delta 1}^{(T)} &=& 4 \pi\alpha_s(m)\,, \qquad
  {\cal V}_{\Delta 1}^{(1)} = 0 \,, \qquad
  {\cal V}_{\Delta 2}^{(T)} = 0 \,, \qquad
  {\cal V}_{\Delta 2}^{(1)} = 0 \,,
\end{eqnarray}
while in Coulomb gauge one gets a different answer
\begin{eqnarray} \label{VdC}
  {\cal V}_{\Delta 1}^{(T)} = 0 \,, \qquad
  {\cal V}_{\Delta 1}^{(1)} = 0 \,, \qquad
  {\cal V}_{\Delta 2}^{(T)} &=& -4 \pi\alpha_s(m)\,, \qquad
  {\cal V}_{\Delta 2}^{(1)} = 0 \,.
\end{eqnarray}
Unlike the on-shell potentials discussed in the previous section, the matching
conditions for off-shell potentials are gauge dependent. Using the expressions
for the soft vertices in Feynman gauge in Appendix~\ref{App_soft}, the one loop
anomalous dimensions are
\begin{eqnarray} \label{convanom}
 && \nu {\partial \over \partial\nu} {\cal V}_{\Delta 2}^{(T)} = -2 \beta_0
  \alpha_s(m \nu)^2 \,, \qquad
  \nu {\partial \over \partial\nu} {\cal V}_{\Delta 1}^{(T)} =
  \nu {\partial \over \partial\nu} {\cal V}_{\Delta 1}^{(1)} =
  \nu {\partial \over \partial\nu} {\cal V}_{\Delta 2}^{(1)} = 0
  \,.
\end{eqnarray}

Including the potential in Eq.~(\ref{V0a}) modifies the matching condition for
the $1/|{\bf k}|$ potentials and also makes this matching gauge dependent.
This is because in the effective theory there are now two new order
$\alpha_s^2/v$ diagrams:
\begin{eqnarray} \label{kconvert}
 \nn \\[-35pt]
 \begin{picture}(75,40)(-5,1)
   \put(9,16){${\cal V}_c$} \put(48,16){${\cal V}_\Delta$}
   \put(50,-1.7){$\Box$}
   \epsfxsize=2.4cm \lower9pt \hbox{\epsfbox{eft2.eps}}
 \end{picture} +
 \begin{picture}(75,40)(-5,1)
   \put(9,16){${\cal V}_\Delta$} \put(48,16){${\cal V}_c$}
   \put(9,-1.7){$\Box$}
   \epsfxsize=2.4cm \lower9pt \hbox{\epsfbox{eft3.eps}}
 \end{picture}
   &=& \frac{i {\cal V}_c^{(T)} {\cal V}_\Delta^{(T)} }{32 m k}\ T^A T^B
      \otimes
      \bar T^A \bar T^B +\frac{i {\cal V}_c^{(T)} {\cal V}_\Delta^{(1)} }
     {32 m k}\ T^A \otimes \bar T^A +\ldots \,. \nn\\
\end{eqnarray}
In Eq.~(\ref{kconvert}) the box labeled by ${\cal V}_\Delta$ denotes an
insertion of both operators in Eq.~(\ref{V0a}), the ellipses denote operators
that vanish by the equations of motion (proportional to $p_1^0-{\bf p^2}/2m$,
$p_3^0-{\bf p^\prime\,^2}/2m$ etc.), and we have defined
\begin{eqnarray}
  {\cal V}_\Delta^{(T)}={\cal V}_{\Delta 1}^{(T)}+{\cal V}_{\Delta 2}^{(T)}\,,
  \qquad
  {\cal V}_\Delta^{(1)}={\cal V}_{\Delta 1}^{(1)}+{\cal V}_{\Delta 2}^{(1)}\,.
\end{eqnarray}
In Feynman gauge the matching conditions in Eq.~(\ref{Lkresult}) now become
\begin{eqnarray} \label{Lkmatch2}
  {\cal V}_{kF}^{(T)} &=& \alpha_s^2(m) \Big( \frac{3 C_A}{4}-\frac{C_d}{4}
  \Big) \,,\qquad
   {\cal V}_{kF}^{(1)} = \alpha_s^2(m) C_1 \,,
\end{eqnarray}
while in Coulomb gauge
\begin{eqnarray} \label{Lkmatch3}
  {\cal V}_{kC}^{(T)} &=& C_A\, \alpha_s^2(m)  \,,
   \qquad\qquad\qquad\quad
   {\cal V}_{kC}^{(1)} = 0 \,. \phantom{zzzzzz}
\end{eqnarray}
The $1/|{\bf k}|$ potential is often referred to as the non-abelian potential.
From Eqs.~(\ref{Lkmatch2}) and (\ref{Lkmatch3}), we see that the $1/|{\bf k}|$
potential only vanishes in QED if  the potential is taken to include off-shell
components and Coulomb gauge is used.

The result in Eq.~(\ref{kconvert}) shows that in the off-shell matching
potential one can make the replacements
\begin{eqnarray}
  V_\Delta^{(T)} & \to & V_\Delta^{(T)} + \zeta^{(T)} \,,  \\
  V_\Delta^{(1)} & \to & V_\Delta^{(1)} + \zeta^{(1)} \,, \nn \\
  V_k^{(T)} & \to & V_k^{(T)} + \frac{1}{32\pi^2} V_c^{(T)} \bigg[ \zeta^{(1)}
     -\frac14 (C_A+C_d) \zeta^{(T)} \Big] \,, \nn \\
  V_k^{(1)} & \to & V_k^{(1)} + \frac{1}{32\pi^2} V_c^{(T)}\: C_1 \zeta^{(T)}
     \,,\nn
\end{eqnarray}
for arbitrary $\zeta^{(T,1)}$.   For instance, at the matching scale taking 
$\zeta^{(1)}=0$, $\zeta^{(T)}=-8\pi\alpha_s(m)$ effectively transforms the 
Feynman gauge result in Eqs.~(\ref{VdF}) and (\ref{Lkmatch2}) into the Coulomb 
gauge result in Eqs.~(\ref{VdC}) and (\ref{Lkmatch3}).  Furthermore, taking 
$\zeta^{(1)}=0$ and $\zeta^{(T)}=4\pi\alpha_s(m)$ transforms the off-shell 
Coulomb gauge result into the on-shell result in section~II. These 
transformations convert terms in the potential that are order $\alpha_s v^0$ 
to order $\alpha_s^2/v$. Similar transformations for the position space color 
singlet potentials have been pointed out previously in 
Refs.~\cite{melnikov,hoang,mm1bram}.

\section{The Quark-Quark Potential}  \label{QQptnl}

The quark-quark potentials in the effective theory can be defined in the same
way as the quark-antiquark potentials. The only difference is that the $V^{(T)}$
terms are now the coefficient of the $T^A \otimes T^A$ tensor, rather than the
$T^A \otimes \bar T^A$ tensor. The computation of the quark-quark potential is
almost identical to that of the quark-antiquark potential. The result can be
obtained from the quark-antiquark potential by omitting the annihilation terms,
and making the substitutions $C_d \to - C_d$ and $\bar T^A \to T^A$. The change
in sign of $C_d$ arises from the identities in Eq.~(\ref{TTiden}). Replacing
$\bar T$ by $T$ in these equations changes the sign of the $C_d$ terms.

The potential in the symmetric and antisymmetric
$QQ$ color channels (the $\bf 6$ and $\bf \bar 3$ for $SU(3)$) are given by
\begin{eqnarray} \label{QQProj}
 \left[\begin{array}{c} V_{\rm symmetric} \cr V_{\rm antisymmetric}
  \end{array}\right]
 =\left[\begin{array}{ccc} 1 &  & {N_c-1\over 2N_c}  \cr
    1 &  & -{N_c+1\over 2N_c}\cr
 \end{array}\right]
 \left[\begin{array}{c} V_{1\otimes 1} \cr V_{T\otimes T} \end{array}
 \right]\,.
\end{eqnarray}
The spin-$1$ $QQ$ combination is spin-symmetric and the spin-$0$  $QQ$
combination is spin-antisymmetric.  For identical fermions in the
initial state, for the symmetric spin-color states we must antisymmetrize the
potential in  the momenta, $V=V({\bf p},{\bf p'}) - V(-{\bf p},{\bf p'})$, and
for the  antisymmetric spin-color states we must symmetrize the potential in
the momenta, $V=V({\bf p},{\bf p'}) + V(-{\bf p},{\bf p'})$. This corresponds
to including the crossed diagrams in the computation of the potentials.

\section{QED}

It is straightforward to obtain the QED potential from our results.   For
oppositely charged particles of charge $\pm Q$, the QED direct potential is
given by $Q^2(V_{1\otimes 1}- V_{T\otimes \bar T})$, where $V_{1\otimes 1}$ and
$V_{T\otimes \bar T}$ are given by our QCD results with $C_1=C_F=T_F=1$,
$C_A=C_d=0$, and $\alpha_s \to \alpha$. The on-shell $1/|{\bf k}|$ potential
does not vanish in this limit. In the results for the annihilation potentials
at the scale $m$ in Eqs.~(\ref{bc2}) and (\ref{amatch}), explicit factors of
$N_c$ were included in the Fierz transformation, so it is simplest to just
separately list the QED limit of these results:
\begin{eqnarray}
  {\cal V}_{s,a}^{(T)} &=& -(i\pi+2-2\ln 2)\: \alpha^2(m) Q^2 \,, \nn\\*[3pt]
  {\cal V}_{s,a}^{(1)} &=& \pi\, \alpha(m) + \bigg\{ -\frac{44}{9} +
   \frac{n_f}{3}\Big( 2\ln 2 -\frac{5}{3} -i\pi \Big) \bigg\}
   \:\alpha^2(m) Q^2\,,\nn\\[3pt]
  {\cal V}_{2,a}^{(T)} &=& 2 (i\pi+2-2\ln 2)\: \alpha^2(m) Q^2  \,,\nn\\*[3pt]
  {\cal V}_{2,a}^{(1)} &=& 0 \,.
\end{eqnarray}
For $e^+e^-$ we have $n_f=0$ and the terms in our $v^0$ potentials
agree with Ref.~\cite{P3}.

For identical particles with charge $Q$, there is only  the direct potential
contribution, which is given by $Q^2(V_{1\otimes 1}+V_{T\otimes T})$, with
$C_1=C_F=T_F=1$, $C_A=C_d=0$, and $\alpha_s \to \alpha$ as above. As discussed
in Section~\ref{QQptnl},  including the crossed diagrams gives a final
potential that is symmetric or antisymmetric in the momenta depending on the
symmetry of the spin state.

\section{Conclusion}

We have computed the $Q \bar Q$ and $Q Q$ scattering amplitudes in QCD to order
$v^2$, and compared our results with previous calculations.  We have also
computed the scattering graphs in vNRQCD, and computed the matching condition
at $\mu=m$ between the two theories. The matching  potential was computed using
on-shell matching, omitting terms which vanish by the equations of motion.  The
result was compared with approaches that include terms in the potential that
vanish on-shell. The one-loop matching coefficients are summarized in
Table~\ref{tab:results}, and can be combined with the two-loop running to give
the potential at next to leading log order. The computation of the heavy quark
production current at next to leading logarithmic order uses these results and
will be discussed in a subsequent publication~\cite{amis3}.

We would like to thank A.~Hoang and J.~Soto for discussions. This work was
supported in part by the Department of Energy under grant DOE-FG03-97ER40546.

\begin{table} {\tighten \caption{ Matching coefficients for the quark-antiquark
potential at $\mu=m$, $\nu=1$. The tree-level contributions are the order
$\alpha_s$ terms, and the one-loop corrections are the order $\alpha_s^2$
terms. The values are for the on-shell potential, so all off-shell potentials
such as the $V_\Delta$ potentials in Eq.~(\ref{V0a}) are set to zero.
Contributions to the matching from annihilation diagrams are given separately
and are denoted by an extra subscript $a$.
\label{tab:results}} }
\[
\begin{array}{lll}
\hline\hline
 & {\cal V}_c^{(T)}\qquad & 4 \pi \alpha_s(m)\\
 & {\cal V}_c^{(1)} & 0\\
 & {\cal V}_k^{(T)} & \alpha_s^2(m) \Big( \frac{7}{8}C_A-\frac{1}{8}C_d \Big)\\
 &  {\cal V}_k^{(1)} & \alpha_s^2(m)\: \frac{1}{2}\,C_1 \\
 & {\cal V}_r^{(T)}  & 4 \pi \alpha_s(m) \\
 & {\cal V}_r^{(1)}  & 0 \\
 & {\cal V}_\Lambda^{(T)} & -6 \pi \alpha_s(m)  -4(C_F+C_A) \alpha_s^2(m) \\
 & {\cal V}_\Lambda^{(1)} & 0 \\
 & {\cal V}_s^{(T)} & -\frac{4}{3} \pi \alpha_s(m)
   +\Big( \frac{1}{2}C_d-\frac{5}{6} C_A-\frac{4}{3} C_F\Big) \alpha_s^2(m)\\
 &  {\cal V}_s^{(1)} & -2 C_1\: \alpha_s^2(m) \\
 & {\cal V}_t^{(T)} & -\frac{1}{3}\pi \alpha_s(m) -\frac{1}{3}(C_F+C_A)
     \alpha_s^2(m) \\
 & {\cal V}_t^{(1)} & 0 \\
 & {\cal V}_2^{(T)} &  \Big( 2 C_F -
    \frac{11}{12} C_d-\frac{2}{3} C_A + \frac{4 }{15}T_F \Big) \alpha_s^2(m) \\
 & {\cal V}_2^{(1)} & \frac{11}{3}\: C_1\: \alpha_s^2(m) \\[3pt]
 \hline
 & {\cal V}_{s,a}^{(T)} & \frac{1}{N_c} \pi \alpha_s(m) + \Big( \frac{1}
   {4N_c}C_d -2 C_1 \Big) (i\pi+2-2\ln 2)\: \alpha_s^2(m) \\
 & &+  \frac{1}{N_c} \bigg\{ \frac{109}{36} C_A-4 C_F  + \frac{n_f T_F}{3}
   \Big( 2\ln 2 -\frac{5}{3} -i\pi \Big) -\frac{8 T_F}{9} \bigg\}\:
   \alpha_s^2(m) \\[5pt]
 & {\cal V}_{s,a}^{(1)} & {(N_c^2-1)\over 2N_c^2}\: \pi\, \alpha_s(m) +
   \bigg\{ \frac{1}{N_c}C_1 +\frac{(N_c^2-1)}{8 N_c^2}C_d \bigg\}
   (i\pi+2-2\ln 2)\: \alpha_s^2(m) \\
 & &+ \frac{(N_c^2-1)}{2N_c^2} \bigg\{ \frac{109}{36} C_A-4 C_F  +
   \frac{n_f T_F}{3}\Big( 2\ln 2 -\frac{5}{3} -i\pi \Big) -\frac{8 T_F}{9}
   \bigg\}\: \alpha_s^2(m) \\[5pt]
 & {\cal V}_{2,a}^{(T)} & -2 \Big( \frac{1}{4N_c}C_d -2 C_1 \Big)
   (i\pi+2-2\ln 2)\: \alpha_s^2(m)  \\[5pt]
 & {\cal V}_{2,a}^{(1)} & -2 \bigg\{ \frac{1}{N_c}C_1 +\frac{(N_c^2-1)}
   {8 N_c^2}C_d \bigg\} (i\pi+2-2\ln 2)\: \alpha_s^2(m) \\[5pt]
   \hline\hline
\end{array}
\]
\end{table}

\newpage

\appendix
\section{Coefficients for the soft Lagrangian}
\label{App_soft}

The coefficient functions for the soft Lagrangian in Eq.~(\ref{Lsoft}) in
Feynman gauge are:
\begin{eqnarray}  \label{nlsoft}
 U^{(0)}_{00} &=&  \frac{1}{q^0}\,,\quad
 U^{(0)}_{0i}  = -\frac{{\mathbf (\two p'-\two p-q)}^i}{\ppp}\,, \quad
 U^{(0)}_{i0}  = -\frac{{\mathbf (p-p'-q)}^i}{\ppp}\,, \quad
 U^{(0)}_{ij}  = \frac{-2 q^0 \delta^{ij} }{\ppp} \nn \,,\\[5pt]
 U^{(1)}_{00} &=& \frac{ {\mathbf (p'+p)\cdot q}}{2m (q^0)^2} -
    \frac{ {\mathbf (p'+p)\cdot q}}{m \ppp} -\frac{i c_F\, \bsigma\cdot
    [{\mathbf q\times (p-p')}] }{m \ppp} +{ ({\bf p'}\,^2-{\bf p}^2) \over
    2 m \ppp } \,, \\[5pt]
 U^{(1)}_{0i} &=& -\frac{{\mathbf (p+p')}^i}{2m q^0} + \frac{i c_F
   ({\mathbf q} \times\bsigma)^i}{2 m q^0} + \frac{q^0 {\mathbf (p+p')}^i}
   {2m \ppp} + \frac{i c_F\, q^0 [{\mathbf (p-p')\times\bsigma}]^i}{2m\ppp} \,,
   \nn \\[5pt]
 U^{(1)}_{i0} &=& -\frac{{\mathbf (p+p')}^i}{2m q^0} - \frac{i c_F [ { \mathbf
  (p-p'+q)\times \bsigma }]^i }{2 m q^0} + \frac{q^0{\mathbf (p+p')}^i}{2m \ppp}
  + \frac{i c_F\, q^0 {[\mathbf (p-p')\times \bsigma}]^i}{2m\ppp} \,,
  \nn \\[5pt]
 U^{(1)}_{ij} &=& \frac{i c_F\, \epsilon^{ijk}\bsigma^k}{2m} + [ 2 \delta^{ij}
  {\mathbf q}^m \!+\! \delta^{im} ( {\mathbf \two p'\!-\! \two p\!-\!q)}^j
  \!+\! \delta^{jm} {\mathbf (p\!-\!p'\!-\!q)}^i]  \nn\\*
 && \times   \Big[ \frac{{\mathbf
  (p\!+\!p')}^m\!+\! i c_F\, \epsilon^{mkl} {\mathbf (p\!-\!p')}^k \bsigma^l }
  {2m ({\mathbf p'-p})^2}
  \Big] \,, \nn\\[5pt]
 U^{(2)}_{00} &=& -\frac{c_D \ppp}{8 m^2 q^0} + \frac{c_S\,\spp}{4 m^2 q^0} +
  \frac{({\bf p\cdot q})^2+({\bf p'\cdot q})^2}{2m^2(q^0)^3}+\frac{(2-c_D)
  {\mathbf (p-p')\cdot q}}{4m^2 q^0} \nn\\
  && +\frac{ (1-c_D) {\mathbf q}^2}{4 m^2 q^0} \,,\nn \\[5pt]
 U^{(2)}_{0i} &=& - \frac{[ {\mathbf p\cdot q}\,{\mathbf (\two p+q)}^i
  +{\mathbf p'\cdot q}\: {\mathbf (\two p'-q)}^i\, ]}{ 4 m^2 (q^0)^2 }
  + \frac{ ic_F [{\mathbf q \times\bsigma}]^i\:
  {\mathbf (p+p')\cdot q} }{ 4 m^2 (q^0)^2 } \nn\\[5pt]
  && \ +\frac{(c_D-1)({\mathbf p-p'+q})^i}{4 m^2} + \frac{ {\mathbf
   (\two p'-\two p-q)}^i }{ 4 m^2 } \bigg[ \frac{c_D}{2}
   -  \frac{c_S\, \spp}{\ppp} \bigg] \nn\\
  && + \frac{ ({\bf p'}\,^2-{\bf p}^2) }{ 4 m^2 ({\bf p'-p})^2 } \bigg[
   ({\bf p}+{\bf p'})^i + c_F\, \spp \bigg] - \frac{{\mathbf (\two p'-
   \two p-q)}^i\,({\bf p'}\,^2-{\bf p}^2)^2 }{4 m^2 ({\bf p'-p})^4} \,,
    \nn\\[5pt]
 U^{(2)}_{i0} &=& - \frac{ [{\mathbf p\cdot q}\,{\mathbf (p+p'+q)}^i +
  {\mathbf p'\cdot q}\: {\mathbf (p+p'-q)}^i\, ] }{ 4 m^2 (q^0)^2 }
  - \frac{ ic_F [{\mathbf (p-p'+q)\times  \bsigma}]^i\:
  {\mathbf (p+p')\cdot q} }{ 4 m^2 (q^0)^2 }   \nn \\[5pt]
  && \ +\frac{(c_D-1) {\mathbf q}^i}{4 m^2} + \frac{ {\mathbf (p-p'-q)}^i }
  { 4 m^2 } \bigg[ \frac{c_D}{2} - \frac{c_S\, \spp}{\ppp} \bigg]  \nn\\
  && - \frac{({\bf p'}\,^2-{\bf p}^2) }{ 2 m^2 ({\bf p'-p})^2 } \bigg[
   ({\bf p}+{\bf p'})^i + c_F\, \spp \bigg] - \frac{{\mathbf (p-
   p'-q)}^i\,({\bf p'}\,^2-{\bf p}^2)^2 }{4 m^2 ({\bf p'-p})^4}
   \,, \nn\\[5pt]
 U^{(2)}_{ij} &=& \frac{{\mathbf (p+p')}^i {\mathbf (p+p')}^j}{4 m^2 q^0} +
  \frac{ c_F^2 {\mathbf (p-p') \cdot q}\: \delta^{ij}}{4 m^2 q^0}
  + \frac{ i c_F {\mathbf (p+p')}^j\: [ {\mathbf (p-p')
  \times \bsigma }]^i }{ 4 m^2 q^0 } \nn\\[5pt]
  && \ - \frac{ i c_F\, \epsilon^{ijk} {\mathbf q}^k
  {\mathbf \bsigma \cdot (p+p') }}{4 m^2 q^0} +  \frac{ i c_F\, \epsilon^{ijk}
  \bsigma^k {\mathbf (p+p')\cdot q }}
  {4 m^2 q^0} + \frac{(1-c_F^2) {\mathbf q}^i({\mathbf p-p'+q})^j}{4 m^2 q^0}
  \nn\\[5pt]
 && \ + \frac{ c_F^2\: {\mathbf q}^2 \delta^{ij} }{4 m^2 q^0}
  - \frac{ i \delta^{ij} q^0 c_S\, \spp}{2 m^2 \ppp} - \frac{2 q^0 \delta^{ij}
  ({\bf p'}\,^2-{\bf p}^2)^2 }{ 4 m^2 ({\bf p'-p})^4 } \,, \nn \\
 W^{(0)}_{\mu\nu} &=& 0 \,,\nn\\
 W^{(1)}_{00} &=& \frac{1}{2m} +\frac{{\mathbf (p-p')\cdot q}}{2m (q^0)^2}
   \,,\quad
 W^{(1)}_{0i}  = -\frac{{\mathbf (p-p'+q)}^i}{2m q^0} \,,\quad
 W^{(1)}_{i0}  = \frac{-{\mathbf q}^i}{2m q^0} \,,\quad
 W^{(1)}_{ij}  = \frac{\delta^{ij}}{2m} \,, \nn \\[5pt]
 Y^{(0)} &=&   \frac{-q^0}{\ppp} \,,\quad
 Y^{(1)} = \frac{ {\mathbf q \cdot (p+p')} + ic_F\, {\mathbf \bsigma \cdot
    [q\times (p-p')] } }{2m \ppp }  \,, \quad \nn\\[5pt]
 Y^{(2)} &=& \frac{ c_D\,q^0 }{8m^2} - \frac{c_S\, \spp q^0 }{4 m^2 \ppp}
    \,,\nn \\
 Z^{(0)}_0 &=& \frac{1}{\ppp} \,,\quad  Z^{(0)}_i =0 \,, \quad
 Z^{(1)}_0 = 0 \,, \quad
 Z^{(1)}_i =  \frac{ -({\mathbf p+p'})^i - i c_F [({\mathbf p-p')\times
 \bsigma]}^i}
   {2m \ppp } \,,\quad \nn\\
 Z^{(2)}_0 &=& -\frac{1}{4 m^2} \bigg[ \frac{c_D}{2} - \frac{c_S\, \spp}
   {\ppp} \bigg] \,,\quad
 Z^{(2)}_i = 0 \,.  \nn
\end{eqnarray}
These expressions differ from those in Ref.~\cite{amis} by terms proportional
to ${\bf p'\,}^2-{\bf p}^2$, and the values in Eq.~(\ref{nlsoft}) are the
complete on-shell expressions. The ${\bf p'\,}^2-{\bf p}^2$ terms were not
needed in calculating the one-loop running of the on-shell order $v^0$
potentials in Ref.~\cite{amis}. In section~III, these terms were used to
compute the running of the off-shell potential in Eq.~(\ref{convanom}), and
this result depends on the fact that we used the on-shell soft Lagrangian. The
coefficients in Eq.~(\ref{nlsoft}) can be written in a manifestly Hermitian
form, however we have instead used ${\bf q'}={\bf q+p-p'}$ and
$q'\,^0=q^0+p^0-p'\,^0$ to eliminate $q\,'$ since this form is more convenient
for calculations. Reparameterization invariance~\cite{repar} can be used to
eliminate $c_S$ by the relation $c_S=2c_F-1$. The running of $c_D$ and $c_F$
are given in Refs.~\cite{run}.

\section{Integrals} \label{App_int}

The effective theory integrals that appear in Eqs.~(\ref{eft1}) and
(\ref{eftVcV0}) are
\begin{eqnarray}
  I_0 &=& \int d^3 q {1\over [({\mathbf q-p})^2+\lambda^2]
  [({\mathbf q-p'})^2 +\lambda^2] [{\bf q^2}-{\bf p}^2-i\epsilon] }
   = \frac{-i}{8\pi |{\mathbf p}| {\mathbf k}^2 } \ln\bigg(
    \frac{\lambda^2}{{\mathbf k}^2} \bigg) \,, \\
  I_F &=& \int d^3 q {1\over ({\mathbf q-p})^2 ({\mathbf q-p'})^2 }
   = \frac{1}{8 |{\mathbf k}| } \nn \,, \\
  I_P &=& \int d^3 q {1\over [({\mathbf q-p'})^2 +\lambda^2]
    [{\bf q^2}-{\bf p}^2-i\epsilon] }
   = \frac{1}{16 |{\mathbf p}| } + \frac{i}{8\pi |{\mathbf p}| }
    \ln\bigg(\frac{2|{\mathbf p}|}{\lambda} \bigg)\,, \nn
\end{eqnarray}
and
\begin{eqnarray}
 && \int d^3 q {{\bf q}^i \over [({\mathbf q-p})^2+\lambda^2]
  [({\mathbf q-p'})^2 +\lambda^2] [{\bf q^2}-{\bf p}^2-i\epsilon] }
  = ({\bf p'+p})^i \: I_A \,, \\
 && \int d^3 q {{\bf q}^i {\bf q}^j \over [({\mathbf q-p})^2+\lambda^2]
  [({\mathbf q-p'})^2 +\lambda^2] [{\bf q^2}-{\bf p}^2-i\epsilon] } =
  \delta^{ij} I_B + ({\bf p'-p})^i({\bf p'-p})^j I_C \nn\\
 &&\qquad\qquad\qquad\qquad\qquad\qquad\qquad\qquad\qquad\qquad
   \qquad\qquad
 + ({\bf p'+p})^i({\bf p'+p})^j I_D \,, \nn
\end{eqnarray}
where ${\bf k}={\bf p'-p}$ and
\begin{eqnarray}
  I_A &=& \frac{ (2\bmag{p}\bmag{k}-{\mathbf k}^2) \pi -2 i {\mathbf k}^2
   \ln\Big(\frac{2\bmag{p}}{\lambda}\Big) -4i{\mathbf p}^2 \ln\Big(
   \frac{\lambda^2}{{\mathbf k}^2}\Big) }{16\pi \bmag{p}
    {\mathbf k}^2 (4{\mathbf p}^2-{\mathbf k}^2) } \,, \\[5pt]
  I_B &=& \frac{ (2\bmag{p}-\bmag{k}) \pi +2 i \bmag{p}
   \ln\Big(\frac{4{\mathbf p}^2}{{\mathbf k}^2}\Big) }{16\pi
   (4{\mathbf p}^2-{\mathbf k}^2) } \,, \nn\\[5pt]
  I_C &=& \frac{ (2\bmag{p}\bmag{k}-{\mathbf k}^2) \pi -2 i (4{\mathbf p}^2
   -{\mathbf k}^2 )-2i {\mathbf k}^2 \ln\Big(\frac{2\bmag{p}}{\lambda}\Big)
   -4i{\mathbf p}^2 \ln\Big( \frac{\lambda^2}{{\mathbf k}^2}\Big) }
    {32\pi \bmag{p} {\mathbf k}^2 (4{\mathbf p}^2-{\mathbf k}^2) } \,,
    \nn\\[5pt]
  I_D &=& \frac{1}{32\pi \bmag{p} {\mathbf k}^2
   (4{\mathbf p}^2-{\mathbf k}^2)^2 } \: \Bigg[ (2\bmag{p}-\bmag{k})^2
   (4\bmag{p}+\bmag{k})\bmag{k} \pi +2 i {\mathbf k}^2 (4{\mathbf p}^2 -
  {\mathbf k}^2 )\nn\\[5pt]
  && \quad -2i {\mathbf k}^2 (12{\mathbf p}^2-{\mathbf k}^2)
  \ln\Big(\frac{2\bmag{p}}{\lambda}\Big)-4i{\mathbf p}^2 (4{\mathbf p}^2
  +{\mathbf k}^2) \ln\Big(\frac{\lambda^2}{{\mathbf k}^2}\Big)  \Bigg]
  \,. \nn
\end{eqnarray}

{\tighten

} 


\begin{references}

\bibitem{Peter} M.~Peter, Phys. Rev. Lett. {\bf 78}, 602 (1997).

\bibitem{Schroder} Y.~Schr\"oder, Phys. Lett. {\bf B447} 321 (1999).

\bibitem{adm}
T.~Appelquist, M.~Dine and I.~Muzinich, Phys. Lett. {\bf B69}, 231, (1977);
T.~Appelquist, M.~Dine and I.~Muzinich, Phys. Rev. {\bf D17}, 2074, (1978).

\bibitem{Caswell}
W.E.~Caswell and G.P.~Lepage,
Phys. Lett. {\bf 167B}, 437 (1986).

\bibitem{BBL}
G.T.~Bodwin, E.~Braaten and G.P.~Lepage,
Phys. Rev. {\bf D51}, 1125 (1995), Erratum ibid. {\bf D55}, 5853
(1997).

\bibitem{Labelle}
P.~Labelle,
Phys.\ Rev.\ {\bf D58}, 093013 (1998).

\bibitem{lm}
M.~Luke and A.V.~Manohar,
Phys. Rev. {\bf D55}, 4129 (1997).

\bibitem{Manohar}
A.V.~Manohar,
Phys. Rev. {\bf D56}, 230 (1997).

\bibitem{gr}
B.~Grinstein and I.Z.~Rothstein,
Phys. Rev. {\bf D57}, 78 (1998).

\bibitem{ls}
M.~Luke and M.J.~Savage,
Phys. Rev. {\bf D57}, 413 (1998).

\bibitem{Pineda}
A.~Pineda and J.~Soto,
Nucl. Phys. Proc. Suppl. {\bf 64}, 428 (1998);


\bibitem{P2}
A.~Pineda and J.~Soto,
Phys.\ Rev.\ {\bf D58}, 114011 (1998).

\bibitem{Beneke}
M.~Beneke and V.A.~Smirnov,
Nucl. Phys. {\bf B522}, 321 (1998).

\bibitem{Gries}
H.W.~Griesshammer,
Phys. Rev. {\bf D58}, 094027 (1998).

\bibitem{P3} A.~Pineda and J.~Soto, Phys. Rev. {\bf D59} 016005 (1999).

\bibitem{LMR} M.E.~Luke, A.V.~Manohar, and I.Z.~Rothstein, hep-ph/9910209.

\bibitem{psIR} N.~Brambilla, A.~Pineda, J.~Soto, and A.~Vairo, Phys. Rev. {\bf
D60}, 091502 (1999).

\bibitem{kp} B.~A.~Kniehl and A.~A.~Penin,
Nucl.\ Phys.\  {\bf B563}, 200 (1999);
B.~A.~Kniehl and A.~A.~Penin,
hep-ph/9911414.

\bibitem{Brambilla} N.~Brambilla, A.~Pineda, J.~Soto and A.~Vairo,
Phys.\ Lett.\  {\bf B470}, 215 (1999);
Nucl.\ Phys.\  {\bf B566}, 275 (2000).

\bibitem{amis} A.V.~Manohar and I.W.~Stewart, hep-ph/9912226.

\bibitem{prodcurrent} For QED see: A.H.~Hoang,
Phys.\ Rev.\ {\bf D56}, 5851 (1997);
Phys.\ Rev.\ {\bf D56}, 7276 (1997).
For QCD see: A.~Czarnecki and K.~Melnikov,
Phys.\ Rev.\ lett.\ {\bf 80}, 2531 (1998);
M.~Beneke, A.~Signer and V.A.~Smirnov,
Phys.\ Rev.\ lett.\ {\bf 80}, 2535 (1998).

\bibitem{Gupta} S.N.~Gupta and S.F.~Radford, Phys. Rev. {\bf D24}, 2309 (1981);
ibid. {\bf D25}, 3430 (1982).

\bibitem{Yndurain}
S.~Titard and F.J.~Yndurain,
Phys.\ Rev.\ {\bf D49}, 6007 (1994).

\bibitem{Georgi}
H.~Georgi,
Nucl.\ Phys.\  {\bf B361}, 339 (1991).

\bibitem{melnikov} K.~Melnikov and A.~Yelkhovsky,
Nucl.\ Phys.\  {\bf B528}, 59 (1998).

\bibitem{hoang} A.~H.~Hoang,
Phys.\ Rev.\  {\bf D59}, 014039 (1999).

\bibitem{mm1bram} N.~Brambilla, A.~Pineda, J.~Soto and A.~Vairo,
hep-ph/0002250.

\bibitem{hasebe} K.~Hasebe and Y.~Sumino, hep-ph/9910424.

\bibitem{faustov} N.~Van-Hieu and R.~Faustov, Nucl. Phys. {\bf 53} 337 (1964).

\bibitem{chen} Y.-Q.~Chen, Y.-P.~Kuang and R.J.~Oakes, Phys.\ Rev. {\bf D52},
264 (1995).

\bibitem{BSS}
M.~Beneke, A.~Signer and V.~A.~Smirnov,
Phys.\ Lett.\  {\bf B454}, 137 (1999).

\bibitem{Redhead} M.L.G.~Redhead, Proc. Roy. Soc. {\bf A220} 219 (1953).

\bibitem{Nieuw} P.V.~Nieuwenhuizen, Nucl. Phys. {\bf B28} 429 (1971).

\bibitem{PTN} W.~Buchmuller, Y.~Ng, and S.~Tye, Phys. Rev. {\bf D24},
3003 (1981); J.~Pantaleone, S.~Tye, and Y.~Ng, Phys. Rev. {\bf D33} 777 (1986).

\bibitem{Andrep}
A.~H.~Hoang,
Phys.\ Rev.\  {\bf D56}, 7276 (1997).

\bibitem{Buras} A.~J.~Buras and P.~H.~Weisz,
Nucl.\ Phys.\  {\bf B333}, 66 (1990).

\bibitem{Dugan} M.~J.~Dugan and B.~Grinstein,
Phys.\ Lett.\  {\bf B256}, 239 (1991).

\bibitem{Herrlich} S.~Herrlich and U.~Nierste,
Nucl.\ Phys.\  {\bf B455}, 39 (1995).

\bibitem{amis3} A.~Manohar and I.~Stewart, UCSD/PTH 00-05.

\bibitem{repar} M.~Luke and A.V.~Manohar,
Phys. Lett. {\bf B286}, 348 (1992).

\bibitem{run} E. Eichten and B. Hill, Phys. Lett. {\rm B243} 427 (1990),
A. Falk, B. Grinstein, and M. Luke, Nucl. Phys. {\rm B357} 185 (1991),
B. Blok, J. K\"{o}rner, D. Pirjol, and J. Rojas, Nucl. Phys. {\rm B496}
358 (1997), C. Bauer and A. Manohar, Phys. Rev. {\rm D57} 337 (1998).


\end{references}
\end{document}